%% file: pap.tex
\documentclass[12pt]{article}
\input{header}

\usepackage{epsfig,sint,macros,macros_static} 
\begin{document}
\input title.tex
\input intro.tex
\input impro.tex
\input tc.tex
\input r0.tex
\input glue.tex
\input disc.tex

\section*{Aknowledgment}
I would like to thank Rainer Sommer, Peter Weisz, Martin L\"uscher, Urs Wenger, Francesco Knechtli and Stephan D\"urr for useful discussions and suggestions. 
Part of the computation was performed on a CRAY-T3e at Forschungszentrum J\"ulich; we thank NIC for providing CPU resources and assistance.

\appendix  
\input app_trans.tex
\input results.tex

\newpage

\input pap.bbl
\end{document}

%% file: header.tex
\usepackage[german,english]{babel}
\usepackage{amsmath}
\usepackage{amssymb}

\usepackage{cite}
\usepackage{latexsym}
\usepackage{graphics}
\usepackage{epsfig}
\usepackage{rotating}
\usepackage{latexsym}
\usepackage{amssymb}
\usepackage{amsfonts}

%% file: title.tex
\begin{titlepage}
\begin{flushright}
  DESY 03-120 \\
\end{flushright}

\vskip 1 cm
\begin{center}
  {\Large\bf Universality and scaling behavior of RG gauge actions }
\end{center}
\vskip 1.0cm
\begin{center}
{\large Silvia Necco}
\vskip 0.8cm
DESY, Platanenallee 6, D-15738 Zeuthen, Germany
\vskip 0.8cm
 necco@ifh.de
\end{center}
\vskip 2.5ex
{\bf{ Abstract}}
\vskip 0.7ex

We study universality and scaling properties of RG gauge actions (Iwasaki and DBW2). In the first part we consider the critical temperature $T_{c}$ and compute the reference energy scale $\rnod$ for critical couplings $\beta_{c}$ corresponding to $N_{t}=3,4,6,8$. The universality of $T_{c}\rnod$ between Iwasaki and Wilson action is confirmed and the scaling behavior of the Iwasaki action is found to be better than the one for the Wilson action. The results for the DBW2 action show larger lattice artefacts. A continuum value $T_{c}\rnod=0.7498(50)$ is extracted. 
We compute also the glueball masses for the states $0^{++}$ and $2^{++}$, investigate the scaling of $m_{0^{++}}\rnod$ and $m_{2^{++}}\rnod$ and point out practical problems which are due to the violation of positivity present in the RG actions.

  \vfill

\eject

\end{titlepage}


%% file: intro.tex
\section{Introduction}
In view of the next unquenched lattice simulations, big efforts were devoted
in the last years to improve both fermionic and gauge actions.\\
Although the lattice artefacts for the standard gauge action start at ${\rm
  O}(a^2)$ and hence one expects that these are somehow less relevant than the ones induced by the fermionic part, it turns out that the gauge term plays an important r\^ole,
and the question which is the most convenient gauge action to adopt has been often addressed.\\
The purpose of adopting alternative actions is not only to improve the scaling
behavior; also features related to chiral symmetry were investigated. In particular, RG actions (Iwasaki, DBW2) have been suggested as good candidates to be used in the next simulations on Ginsparg-Wilson/domain wall fermions; interesting characteristics were observed, such as the suppression of small instantons and dislocations and a possible remedy of the problem of residual chiral symmetry breaking for domain wall fermions \cite{Orginos:2001xa,Aoki:2002vt}. Other authors \cite{Jansen:2003jq}, however, pointed out possible problems connected to these actions and proposed an alternative method to reduce residual-mass effects that works very well also with the Wilson plaquette action.\\
This increasing interest and discussion in improved gauge actions motivates more investigations into their properties, starting from the basic ones, like universality and scaling behavior.\\
There are in principle several quantities that one can use to quantify the
lattice artefacts and to test universality by comparing the results with the
plaquette action known in the literature; in particular in this work we will consider the critical deconfining
temperature $T_{c}$ and the glueball masses for the states $0^{++}$ and
$2^{++}$.\\
In a previous work \cite{Okamoto:1999hi}, the scaling of the ratio $T_{c}/\sqrt{\sigma}$ has been investigated and lead for the Iwasaki action to a continuum limit which
is in disagreement with the one obtained with the Wilson action. More likely than a possible violation of universality, this fact is related to the definition of the string tension $\sigma$.\\
We will
perform the same analysis but using the scale $\rnod\approx 0.5\fm$ \cite{pot:r0} instead of $\sigma$. It is well known that the extraction
of the string tension from the static potential at finite quark separations
is problematic and leads to systematic errors, which are difficult to control. It is preferable
to define the scale by using intermediate distance properties.\\
After computing $\rnod/a$ at several lattice spacing for Iwasaki
and DBW2 actions we will perform a continuum extrapolation of the quantity
$T_{c}\rnod$, discuss universality and compare the scaling violations for the
different actions.\\
We will point out possible problems that can occur by adding
irrelevant operators to the plaquette action, as the violation of physical positivity.\\
The second part of the work is dedicated to the evaluation of the glueball
masses. In particular the $0^{++}$ mass is a promising observable to discuss
the lattice artefacts, since for the Wilson action large scaling violations
have been observed.\\
Despite the large errors and the difficulties in the calculations, we will discuss our results, draw indicative conclusions and remark on open questions.


%% file: impro.tex
\section{Improved gauge actions}
In this work we will consider $\SUthree$ Yang-Mills lattice actions formulated on the basis of renormalization group (RG) considerations.
The most popular examples are the Iwasaki \cite{Iwasaki:1983ck} and DBW2 \cite{deForcrand:1997bx,Takaishi:1996xj} actions, which are restricted to a two-parameter space and include planar rectangular  
$(1\times 2)$ loops in addition to the usual plaquette term,
\begin{equation}\label{impr_action}
S=\beta\sum_{x}\left(c_{0}\sum_{\mu<\nu}\left\{1-\frac{1}{3}\Re W_{\mu\nu}^{1\times 1}(x)\right\}+c_{1}\sum_{\mu,\nu}\left\{1-\frac{1}{3}\Re W_{\mu\nu}^{1\times 2}(x)\right\}\right),
\end{equation} 
with the normalization condition $c_{0}=1-8c_{1}$.\\
The coefficient $c_{1}$ in \eq{impr_action} takes different values for various choices of alternative actions
\begin{equation}\label{c1_impr}
c_{1} =\left\{\begin{array}{ll}
-1/12  & \textrm{Symanzik, tree level impr.}\\
-0.331 & \textrm{Iwasaki, RG}\\
-1.4088 & \textrm{DBW2, RG}
\end{array}\right.
\end{equation}
where we included also the ${\rm O}(a^2)$ tree-level improved Symanzik action \cite{Weisz:1983zw},\cite{Weisz:1984bn}.\\
First of all one can notice that the strength of the rectangular loops for RG
improved actions is significantly larger than what is needed in order to cancel the ${\rm O}(a^2)$ effects at tree level. We will later discuss this further.\\ 
We will also consider the gauge FP action \cite{Hasenfratz:1994sp} for comparison, in particular the study on the lattice artefacts performed in \cite{Niedermayer:2000yx}.


%% file: tc.tex
\section{The critical temperature $T_{c}$}
It is well known \cite{Polyakov:1978vu,Susskind:1979up} that pure Yang-Mills theory undergoes a first order phase transition at some finite temperature $T_{c}$. On the lattice,
the critical temperature is determined by evaluating the critical coupling $\beta_{c}$
\begin{equation}
\frac{1}{T_{c}}=N_{t}a(\beta_{c}),
\end{equation}
where $N_{t}$ is the number of lattice points in the time-like direction with periodic boundary conditions. The extension in the space-like directions is supposed to be infinite.\\
There are several methods for determining $\beta_{c}$; for example, one can locate the peak in the Polyakov loop susceptibility.\\
In addition to its intrinsic importance as a fundamental non-perturbative prediction, $T_{c}$ provides also a useful quantity to study the lattice artefacts for different gauge actions and to test universality.\\
For known values of $\beta_{c}$ at given $N_{t}$ for different actions one can
refer to \cite{Boyd:1996bx,Beinlich:1997ia,Okamoto:1999hi,deForcrand:1999bi,Niedermayer:2000yx};
for Wilson, Iwasaki and DBW2 actions, these values are collected in \tab{tab_betacrit}.\\

\begin{table}
\begin{center}
\begin{tabular}{c c c c c}
\hline
        &  $\beta_{c}$ & Wilson \cite{Beinlich:1997ia}  & Iwasaki \cite{Okamoto:1999hi}  & DBW2 \cite{deForcrand:1999bi}  \\
$N_{t}$ &              &                 &                  &               \\
\hline
3       &              &                 & 2.1551(12)       & 0.75696(98) \\  
4       &              & 5.6925(2)       & 2.2879(11)       & 0.82430(95) \\
6       &              & 5.8941(5)       & 2.5206(24)       & 0.9636(25) \\ 
8       &              & 6.0624(12)      & 2.7124(34)       &             \\
12      &              & 6.3380(23)      &                  &               \\
\hline
\end{tabular}
\end{center}
\caption{\footnotesize{The critical coupling for Wilson,
        Iwasaki and DBW2 actions.  \label{tab_betacrit}}}
\end{table}
\begin{table}[h]
\begin{center}
\begin{tabular} {c c c}
\hline
action          & $N_{t}$   & $T_{c}/\sqrt{\sigma}$ \\
\hline
Wilson\cite{Beinlich:1997ia} & $\infty$  &  0.630(5)                     \\
Iwasaki \cite{Okamoto:1999hi} & $\infty$  &  0.651(12)                     \\
DBW2 \cite{deForcrand:1999bi} & $\infty$  &  0.627(12)                     \\
Sym. tree level\cite{Beinlich:1997ia} & $\infty$  &  0.634(8) \\
1 loop tadpole impr. \cite{Bliss:1996wy}  &  $\infty$  &  0.659(8)\\
FP \cite{Niedermayer:2000yx}             & 4         &  0.624(7)                     \\
\hline
\end{tabular}
\end{center}
\caption{\footnotesize{\label{tc_sqrtsigma}Results for the deconfining temperature in units
  of the string tension from different actions. The continuum extrapolations are taken from \protect\cite{glueb:teper98}. }}
\end{table}

\begin{figure}
\begin{center}
\includegraphics[angle=-90,width=9cm]{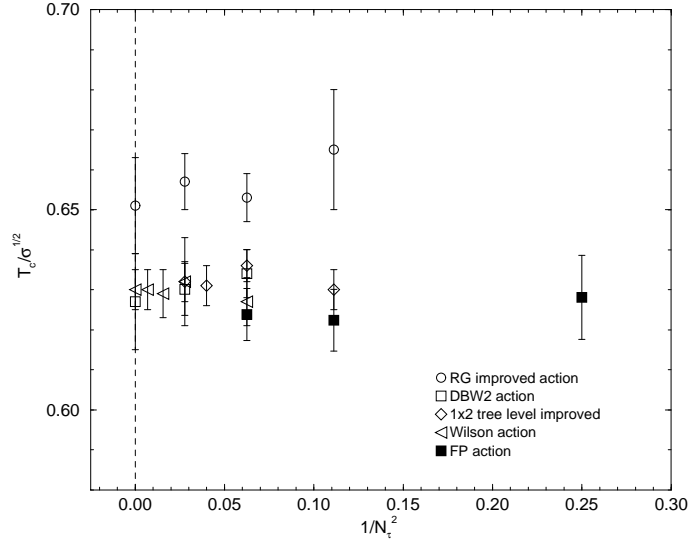}
\end{center}
\caption{\footnotesize{$T_{c}/\sqrt{\sigma}$ as function of $1/N_{t}^2$ for different actions, from \protect\cite{Niedermayer:2000yx}. By ``RG improved'' is meant here the Iwasaki action.}\label{tc_sqrtsigma_fig}}
\end{figure}

The available results in the literature are mostly expressed in terms of the string tension $\sigma$; the results for the quantity $T_{c}/\sqrt{\sigma}$ are reported in \tab{tc_sqrtsigma}, where the continuum extrapolations are from \cite{glueb:teper98}. 
\Fig{tc_sqrtsigma_fig} collects the results for Wilson, Iwasaki, DBW2,
Symanzik tree level and FP action, and was taken from
\cite{Niedermayer:2000yx}, where the latest evaluation of $T_{c}/\sqrt{\sigma}$ (FP action) has been performed.

The first observation from \fig{tc_sqrtsigma_fig} is that for this specific quantity the discretization effects appear not to be very significant and hence is not possible to arrive at precise conclusions about the lattice artefacts for different actions.

Furthermore, one notices that for the Iwasaki and the Wilson actions a difference of order $2\sigma$ in the continuum results is observed. 
The most drastic explanation for this discrepancy could be a violation of 
universality, but this scenario seems unrealistic; a
more natural explanation is that the string tension is difficult to determine
and systematic errors due to this were not included in all calculations.
It is preferable
to use $\rnod$ \cite{pot:r0} to reliably set the scale.\\
Let us recall that the string tension is related to the properties of the force between static quarks for distances $r\rightarrow\infty$, and hence its evaluation at finite $r$ can in principle contain large systematic errors.\\
On the contrary, $\rnod\approx 0.5\fm$ is extracted from the force at
intermediate distances and can be evaluated very precisely. For the Wilson
action this quantity has been evaluated in the coupling range $5.7\leq\beta\leq 6.92$ \cite{pot:r0_SU3,Necco:2001xg}.\\

%% file: r0.tex
\section{Evaluation of $\rnod/a$ for RG actions}\label{ppp}
For the Wilson action, the values of $\rnod/a$ corresponding to the critical couplings at different $N_{t}$ can be easily obtained by the parametrization formula in \cite{pot:r0_SU3}.\\
For the Iwasaki and DBW2 actions there was up to now no precise evaluation of $\rnod/a$ and we performed new numerical simulations with this purpose.\\
In our computation we followed essentially the procedure adopted 
in \cite{pot:r0,pot:r0_SU3} for the plaquette action.
We applied the smearing procedure\cite{smear:ape} to the spatial links and 
the multi-hit method \cite{PPR} to the time-like links for the variance reduction.\\
For each spatial separation
$r$ we constructed Wilson loop correlation matrices
\begin{equation}\label{corrma}
C_{lm}(t)=\Bigg\langle\Tr\left\{V_{l}(0,r\hat{1})\overline{V}(r\hat{1},r\hat{1}+t\hat{0})V_{m}^{\dagger}(t\hat{0},r\hat{1}+t\hat{0})\overline{V}^{\dagger}(0,t\hat{0}) \right\}\Bigg\rangle
\end{equation}
with $l,m=1,...,M$; $V_{l}(x,y)$ indicates the product of link variables connecting $x$ and $y$ at smearing level $l$ in a spatial direction, while $\overline{V}(x,y)$ is the product of time-like links between $x$ and $y$ after the application of the multihit procedure.\\ 
In particular we have chosen $M=4$, and 
the number of smearing iterations for each level has been determined with the same
criterion as for the Wilson action \cite{pot:r0_SU3}.\\
We adopted a hybrid algorithm with $N_{or}$ over-relaxation steps per
heat-bath step \cite{HOR1,HOR2} and increased $N_{or}$ with $\beta$ according to 
\begin{equation}
N_{or}\approx 1.5(\rnod/a).
\end{equation}
The simulation parameters are reported in \tab{simpar_iwasaki}.
For the DBW2 action, besides the three
values of $\beta_c$ known in the literature (\tab{tab_betacrit}), we decided to
evaluate $\rnod/a$ also for a larger $\beta=1.04$, which should roughly
correspond to $\beta=6$ for the Wilson action and has been used in quenched
simulations \cite{Aoki:2002vt}.

\begin{figure}
\begin{center}
\includegraphics[width=8.7cm]{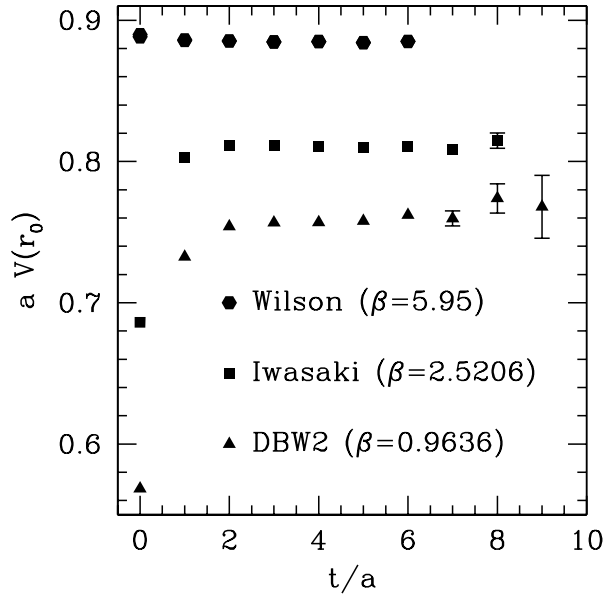}
\vspace{-1cm}
\caption{\footnotesize{The effective potential as function of $t$ for
    $r\approx\rnod$ ($\rnod/a\sim 4-5$) for the Wilson action and for
    the RG-improved Iwasaki and DBW2 action.}\label{f_potr0}}
\end{center}
\end{figure}

We started our analysis by observing the time-dependence of the effective
potential evaluated from the diagonal elements of the correlation matrices.\\
\Fig{f_potr0} shows the effective potential
\begin{equation}\label{v_effic}
a V(r)=-\ln\left(\frac{C_{22}(t)}{C_{22}(t-a)}\right)
\end{equation}
evaluated for $l=m=2$, which corresponds to what was estimated to be the optimal smearing for the Wilson action in \cite{pot:r0_SCRI}.

It has been pointed out in \cite{Luscher:1984is} (see app. \ref{app_trans}) that actions containing terms in addition to the plaquette violate the physical
positivity, and hence one expects also negative contributions to
the spectral decomposition of correlation functions.\\
These negative contributions for the RG actions are 
evident in \fig{f_potr0} and, as expected from our discussion in app. \ref{app_trans}, they are invisible starting from a certain $t$ which
depends on the coefficient $c_{1}$ of \eq{c1_impr}.\\
The most important consequence of this phenomenon regards the applicability of
the variational method \cite{varia:michael,phaseshifts:LW}, which is mathematically founded on the positivity 
of the correlation matrix $C_{lm}(t)$ at a certain small
$t=t_{0}$; this condition is verified only for $t_{0}\gg t_{min}$, but on the
other hand $t_{0}$ can not be arbitrarily large because the statistical
errors increase exponentially with $t$ and make the inversion of $C_{lm}$ 
impracticable.\\
One observes however a quite satisfactory plateau in the effective potential,
starting at sufficiently large $t$. In that sense the violation of physical
positivity does not represent a real trouble for these correlation functions,
but can become quite problematic for example for the extraction of the
glueball masses (see sec. \ref{sect_glueball}).\\
We decided to extract the potential from
\eq{v_effic} at $t/a=(3-4)$ without applying the variational method.
The systematic error was estimated by taking the difference between this value and what one
would obtain extracting the potential at $(t+a)$.
We linearly added systematic and statistical errors; at small $r$ the total uncertainty is dominated by the systematic one, while at large distances the situation is in general reversed.

\subsection{The force at tree level}
\begin{figure}
\begin{center}
\includegraphics[width=9cm]{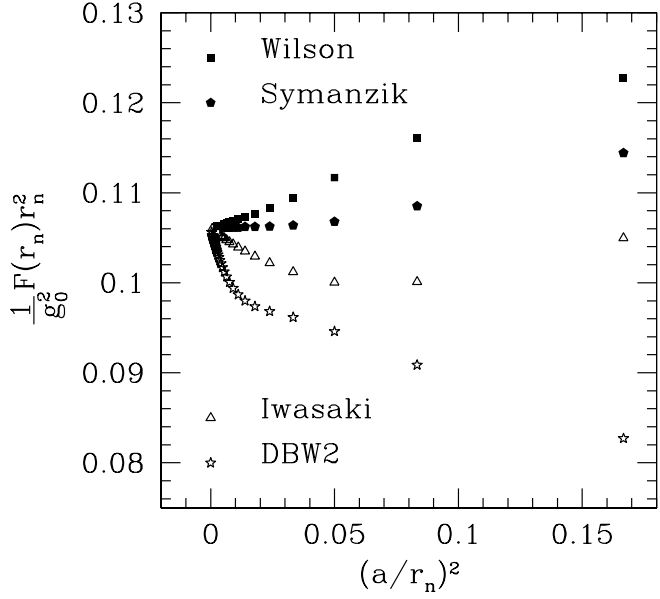}
\end{center}
\vspace{-1.5cm}
\caption{\footnotesize{The force at tree level for several actions.\label{fig_forcetree}}} 
\end{figure}

A tree-level study of the force for the different actions can furnish
important hints on how the continuum limit is approached, in particular in the region of 
small couplings.\\
Explicitly, for the action \eq{impr_action} the tree-level force is given by \cite{Weisz:1983zw} 
\begin{equation}\label{force_treelevel_improved}
F(r^{\prime}) = \frac{V(r)-V(r-a)}{a}=
\end{equation}
$$
                     =
                     -\frac{4}{3}\frac{g_{0}^{2}}{a}\int_{-\pi}^{\pi}\frac{d^{3}k}{(2\pi)^3}\frac{\cos(rk_{1}/a)-\cos((r-a)k_{1}/a)}
                     {4\left(\sum_{j=1}^{3}\sin^{2}(k_{j}/2)-4c_{1}\sum_{j=1}^{3}\sin^{4}(k_{j}/2)\right)}+
                     {\rm O} (g_{0}^4a^2)=
$$
$$
=F_{tree}(r')+{\rm O} (g_{0}^4a^2),
$$
where $g_{0}^2=6/\beta$.\\
In \fig{fig_forcetree} the quantity
$\frac{{r_{n}}^{2}}{g_{0}^2}F_{tree}(r_{n})$, with $r_{n}=r-a/2$ is plotted as function
of $(a/r_{n})^2$ for different actions. 
One can notice that the RG actions, in particular
the DBW2, show at tree level large lattice artefacts at small
$(a/r^{\prime})$. In that sense, at tree level the RG actions are "over-corrected" and
introduce lattice artefacts of the same order or even larger than what is
expected with the usual plaquette action.
This fact indicates that the continuum extrapolation should be considered 
with great care, because unless $(a/r')$ is very small, one can not be sure to be in the region where the leading discretization errors are quadratic in $a$.\\
Beside the naive definition of the force
\begin{equation}\label{force_def}
F(\r')=[V(r)-V(r-a)]/a,
\end{equation}
with $r'=r_{n}$, 
we followed the procedure reported in \cite{pot:r0} by introducing a tree-level
improved definition with $r'=\rI$ such that
\begin{equation} 
F_{tree}(\rI)=\frac{4}{3}\frac{g_{0}^2}{4\pi\rI^2},
\end{equation}
with no lattice artefacts at tree level. For the Wilson action this choice turned out to improve substantially the scaling behavior of the force \cite{pot:r0,pot:r0_SU3}. \\
From \eq{force_treelevel_improved} one obtains
\begin{equation}\label{e_rI_new}
 (4\pi\rI^2)^{-1}=-[G(r,0,0)-G(r-a,0,0)]/a,
\end{equation}
where 
\begin{equation}\label{prop_impr}
  G({\vec r})=\frac{1}{a}\int_{-\pi}^{\pi}\frac{d^{3}k}{(2\pi)^3}\frac{\prod_{j=1}^{3}\cos(r_{j}k_{j}/a)}{4(\sum_{j=1}^{3}\sin^{2}(k_{j}/2)-4c_{1}\sum_{j=1}^{3}\sin^{4}(k_{j}/2))}, 
\end{equation}
is the scalar free lattice propagator associated to the action \eq{impr_action}.
We computed $\rI$ by solving the integral \eq{prop_impr} numerically.

\begin{table}[h]
\begin{center}
Iwasaki action:
\end{center}
\begin{center}
\begin{tabular}{c c c c c}
\hline
$L/a$ & $\beta$  &   $n_l$    & $N_{or}$ & $N_{meas}$  \\
\hline
8       & 2.1551   &  0,2,4,6     & 3       & 20000\\
12      & 2.2879   &  0,4,9,13    & 4       & 4000\\
24      & 2.5206   &  0,12,25,37  & 8       & 645 \\
32      & 2.7124   &  0,18,36,54  & 9       & 370  \\
\hline
\end{tabular}
\end{center}
\begin{center}
DBW2 action :
\end{center}
\begin{center}
\begin{tabular}{c c c c c}
\hline
$L/a$ & $\beta$  &   $n_l$    & $N_{or}$ & $N_{meas}$  \\
\hline
10      & 0.75696  &  0,2,4,6     & 3        & 12000  \\  
12      & 0.8243   &  0,4,9,13    & 4        & 6000\\
16      & 0.9636   &  0,10,20,30  & 8        & 800 \\
24      & 1.04     &  0,18,36,54  & 9        & 220\\
\hline
\end{tabular}
\end{center}
\caption{\footnotesize{Simulation parameters for the Iwasaki and DBW2 actions
    for the evaluation of $\rnod/a$. $L$ is the lattice extension (the spatial
    and temporal extensions are equal), $n_{l}$
    represents the number of smearing iterations for each level.}\label{simpar_iwasaki}}
\end{table}

\subsection{Results}
Our numerical results for the potential and the force at finite lattice
spacing are collected in the tables \ref{tab_iwasaki_results} and
\ref{tab_dbw2_results} in \app{numerical_results}.\\
Once the force has been evaluated, we extracted the value of $r_{0}/a$ by
using a local interpolation formula; the results are
reported in \tab{r0_results}. We adopted both the naive
and the tree-level improved definition of the force.\\
\begin{table}[h]
\begin{center}
Iwasaki action :
\end{center}
\begin{center}
\begin{tabular}{c c c c}
\hline
$\beta$  &  $\rnod/a$ $(r_{n})$  & $\rnod/a$ $(\rI)$ & $\rnod/a$ (\protect\eq{fit_iwasaki}) \\
\hline
2.1551   &   2.311(5)(9)      & 2.320(6)(9) & 2.333                    \\
2.2879   &   3.026(4)(3)      & 3.026(5)(1) & 3.021                  \\
2.5206   &   4.535(6)(4)      & 4.511(8)(1) & 4.514                   \\
2.7124   &   6.020(15)(25)     & 5.999(15)(19) & 5.983                      \\
\hline
\end{tabular}
\end{center}
\begin{center}
DBW2 action :
\end{center}
\begin{center}
\begin{tabular}{c c c}
\hline 
$\beta$  &  $\rnod/a$ $(r_{n})$  & $\rnod/a$ $(\rI)$  \\
\hline
0.75696   &  2.430(5)(20)             &   2.225(4)(11) \\
0.8243    &  3.129(23)(1)             &   3.036(17)(4) \\
0.9636    &  4.606(13)(17)            &   4.556(17)(20)\\
1.04      &  5.500(29)(7)             &   5.452(26)(8) \\
\hline
\end{tabular}
\end{center}
 \caption{\footnotesize Results for $\rnod/a$
   evaluated at different $\beta=\beta_{c}$ for Iwasaki and DBW2 actions, using the naive definition of the
   force or the tree-level improved \protect\eq{e_rI_new}. For the Iwasaki action, the fourth column is obtained from the fits \protect\eq{fit_iwasaki}.\label{r0_results}}
\end{table}
The first error contains the statistical uncertainty summed to the systematic
one due to the interpolation of the force. The second error is the systematic
uncertainty coming from different choices of $t$ in the effective potential
\eq{v_effic}.\\
One can notice that for the
DBW2 action the choice of $r_{n}$ or $\rI$ in the definition of the force
leads to results which can be quite different from each other, above all at small $\rnod/a$; we expected this feature by investigating the force at tree level.
This ambiguity
will make the  discussion of the lattice artefacts difficult, because the
possible conclusions will depend on which definition of the force one has used
and not on intrinsic properties of the action.\\
For the Iwasaki action the results obtained through the two definitions are not significantly different.

\subsection{Parametrization of $\rnod/a$}
\begin{figure}
\begin{center}
\includegraphics[width=8.5cm]{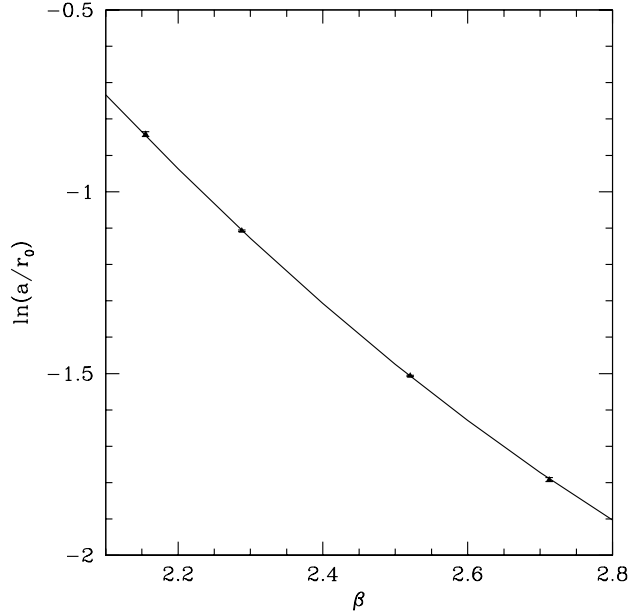}
\end{center}
\vspace{-1cm}
\caption{\footnotesize{Parametrization of $\rnod/a$ for the Iwasaki
    action, \protect\eq{fit_iwasaki}, using the tree-level improved
    definition of the force.}\label{r0_par_iwasaki}}
\end{figure}

Following the strategy of \cite{pot:r0_SU3}, one can attempt a
phenomenological parametrization of $\rnod/a$ in the range of couplings under consideration.\\
For the Iwasaki action the four values of $\rnod$ (obtained by adopting $\rI$)
were fitted in the form
\begin{equation}\label{fit_iwasaki}
\ln(a/\rnod)=c_{1}+c_{2}(\beta-3)+c_3 (\beta-3)^2,
\end{equation}
yielding the numerical results
\begin{equation}
c_{1}= -2.1281   ,\quad c_{2}=-1.0056    ,\quad c_{3} =0.6041        .
\end{equation}
in the range $2.1551\leq\beta \leq 2.7124$.\\
The results and the fit formula are shown in \fig{r0_par_iwasaki}; the
accuracy is about $0.6\%$ at $\beta=2.1551$ and $0.8\%$ at
$\beta=2.7124$. 
\begin{figure}
\begin{center}
\includegraphics[width=8.5cm]{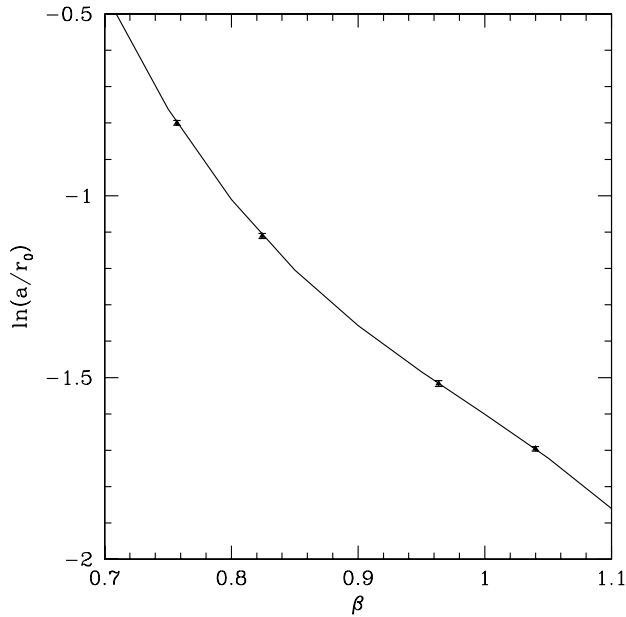}
\end{center}
\vspace{-1cm}
\caption{\footnotesize{Parametrization of $\rnod/a$ for the DBW2
    action, \protect\eq{fit_dbw2}, using the tree-level improved definition of the force.}\label{r0_par_dbw2}}
\end{figure}

For the DBW2 action we had to use a four-parameter representation
\begin{equation}\label{fit_dbw2}
\ln(a/\rnod)=d_{1}+d_{2}(\beta-1)+d_{3}(\beta-1)^2 +d_{4}(\beta-1)^3,
\end{equation}
with
\begin{equation}
d_{1}=- 1.6007,\quad d_{2}=-2.3179,\quad d_{3} =-0.8020,\quad d_{4}=-19.8509 ,
\end{equation}
for the range $0.75696\leq\beta\leq 1.04$, where the results always refer to the tree-level improved definition of the
force. The parametrization is plotted in \fig{r0_par_dbw2}.\\
Note that in \cite{Aoki:2002vt} the value $\rnod/a=5.24(3)$ for the DBW2 action
at $\beta=1.04$ is quoted. Our value differs about $3\%$ from that one.
\section{Scaling of $T_{c}\rnod$}
\begin{figure}
\begin{center}
\includegraphics[width=9cm]{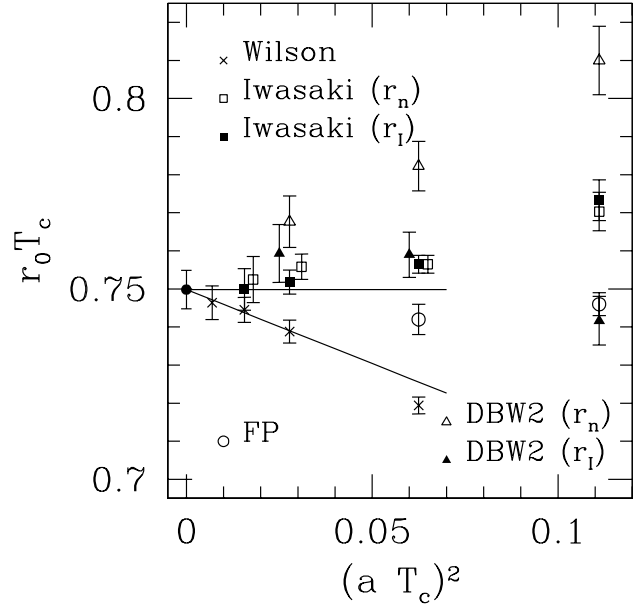}
\end{center}
\vspace{-1cm}
\caption{\footnotesize{$T_{c}r_{0}$ for different actions. The $x$ coordinates were slightly shifted for clarity.}\label{fig_tcr0}}
\end{figure}

Once $\rnod/a$ at the given couplings is known, one can finally consider the
renormalized quantity
\begin{equation}
\rnod T_{c}=\frac{1}{N_{t}}\frac{\rnod}{a}(\beta_c).
\end{equation}
The results for Iwasaki and DBW2 actions are given in \tab{tcr0_results},
together with the values obtained using the Wilson action. Also in this case,
we show both the results obtained with the naive and with the tree-level
improved definition of the force. The error in $T_{c}\rnod$ is the quadratic sum of the
error for $\rnod/a$ and the uncertainty in $\beta_c$, which can be translated into an uncertainty in $\rnod$ by using the parametrization formulas
\eq{fit_iwasaki}, \eq{fit_dbw2}.
In our evaluations the error for $\beta_c$ and the uncertainty in $\rnod$ are
roughly of the same order.\\
We expect that the leading lattice artefacts are of order $a^2$, such that the
continuum limit is approached in the following way
\begin{equation}\label{cont_extr_tc}
T_{c}r_{0}=T_{c}r_{0}|_{a=0}+s\cdot (a T_{c})^{2}+ {\rm O} (aT_c)^4.
\end{equation} 
The results for $T_{c}r_{0}$, together with the continuum extrapolation, are
shown in \fig{fig_tcr0}.
\begin{figure}
\begin{center}
\includegraphics[width=9cm]{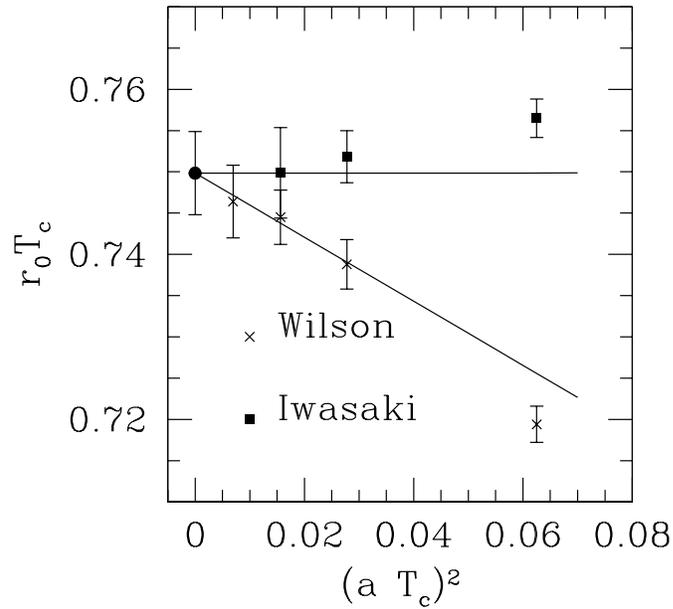}
\end{center}
\vspace{-1cm}
\caption{\footnotesize{Continuum extrapolation of $T_{c}r_{0}$ for the Iwasaki and Wilson action, using the constrained fit \protect\eq{cont_extr_tc}.}\label{fig_tcr0_extr}}
\end{figure}

For the Iwasaki action there is
no appreciable difference between the results obtained with
$r_{n}$ and $r_{I}$ and in both cases the data show better scaling properties in
comparison to the Wilson action. Furthermore, the value obtained at
$N_{t} =8$ is in full agreement with the continuum result evaluated
through the Wilson action and hence the universality is confirmed;
this supports the conclusion that the disagreement observed in
$T_{c}/\sqrt{\sigma}$ is indeed due to the difficulty in evaluating the string
tension, particularly at small lattice spacings, and it is necessary
to set the scale through a more reliable quantity.\\
Also for the DBW2 action the scaling properties are improved,
although only by adopting $r_{I}$ instead of $r_{n}$, so that it is more difficult
to make a statement about the lattice artefacts in this case.\\
A constrained fit of the form \eq{cont_extr_tc}
including the points with $N_{t}\geq 6$
for Iwasaki and Wilson actions yields the continuum result (\fig{fig_tcr0_extr})
\begin{equation}
T_{c}r_{0}=0.7498(50).
\end{equation} 
At $N_{t}=6$ the Wilson action shows scaling violations for $r_{0}T_{c}$ of 
about $1.5\%$, while they are $0.3\%$ for the Iwasaki action.\\
For $N_{t}=4$ the discretization errors for the Wilson action increase to
$4\%$, while for the Iwasaki action they remain very small ($0.6\%$).\\
In \fig{fig_tcr0} we included also the results obtained with the FP action
\cite{Niedermayer:2000yx}, which also show a good scaling within $1\%$ even on coarse lattices
corresponding to $N_t =3,2$. One has however to mention that for those
lattices the determination of $\rnod/a$ contains large systematic
uncertainties, as pointed out by the authors.
\begin{table}
\begin{center}
\begin{tabular}{c c c c c c c }
\hline
      &  $T_{c}\rnod$: &  Wilson  & Iwas.($r_{n}$) & Iwas.($\rI$) & DBW2($r_{n}$) & DBW2($\rI$)\\   
$N_t$ &               &          &           &          &               &    \\
\hline
3    &                &         &    0.7703(50)        &  0.7733(53) & 0.8100(90) & 0.7417(90) \\
4    &                & 0.7194(22)  & 0.7565(23)  & 0.7565(23)       & 0.7822(65)  &  0.7590(56) \\ 
6    &                & 0.7388(30)  & 0.7558(33)  & 0.7518(31)       & 0.7676(68) &  0.7593(76)\\
8    &                & 0.7445(33) & 0.7525(60) & 0.7499(55)  &  & \\
12   &                & 0.7464(44) &   &   & & \\ 
\hline
\end{tabular}
\end{center}
\caption{\footnotesize{Results for $T_{c}\rnod$.}\label{tcr0_results}}
\end{table}
\section{Scaling of $\alphaqqbar(\mu)$}
\begin{figure}
\begin{center}
\includegraphics[width=9cm]{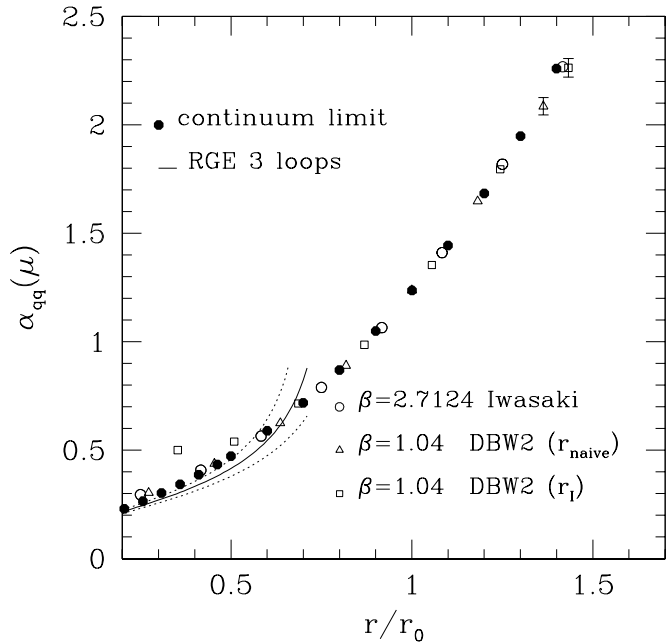}
\end{center}
\vspace{-1cm}
\caption{\footnotesize{$\alpha_{\qqbar}$ at finite lattice spacing for Iwasaki and DBW2 action compared with the continuum result. The solid line represents the 3-loop RG perturbative prediction of the running coupling; the dashed lines correspond to its uncertainty \protect\cite{Necco:2001gh}.}\label{alphaqq_artifacts}}
\end{figure}

\begin{figure}
\begin{center}
\includegraphics[width=8.5cm]{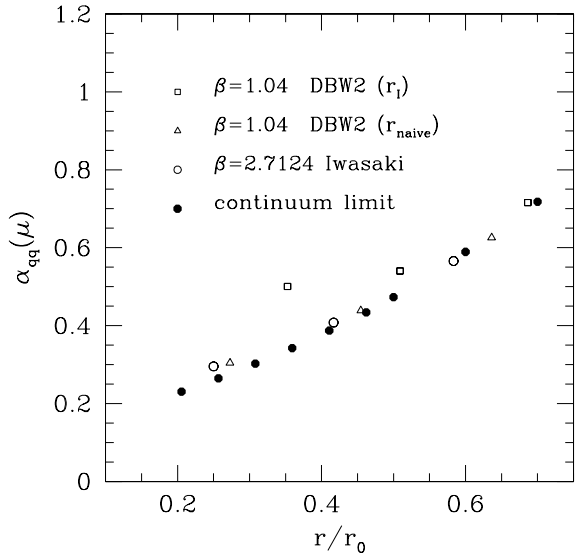}
\end{center}
\vspace{-1cm}
\caption{\footnotesize{$\alpha_{\qqbar}$ at finite lattice spacing for Iwasaki and DBW2 action compared with the continuum results in the short distance region.}\label{alphaqq_artifacts_det}}
\end{figure}

Another interesting observable that can be used to test scaling violations
is the dimensionless coupling $\alphaqqbar(\mu)$ obtained from the force.
For this purpose we compared our present determination of the force at finite lattice spacing  with the results of 
\cite{Necco:2001xg}, where the continuum extrapolation has been performed in the region $0.05\fm\leq r\leq 0.8\fm$.\\
We point out that we only determined the on-axis potential and hence we can not
investigate violations of rotational invariance which would require the
evaluation of off-axis quantities.\\
The coupling $\alphaqqbar$ is defined in terms of the force by the simple relation
\begin{equation}
\alphaqqbar(\mu)=\frac{3}{4}F(r)r^2,\quad \mu=\frac{1}{r}.
\end{equation}
\Fig{alphaqq_artifacts} shows $\alphaqqbar$ in the continuum limit
and the
results obtained with the Iwasaki and DBW2 actions at the largest $\beta$ at our disposal. For the Iwasaki action no appreciable difference in the results obtained with $r_{n}$ and $\rI$ can be seen, while for the DBW2 action the discrepancy becomes large at small distances.\\
At large enough distances one obtains a good scaling in the
coupling, and one does not observe scaling violation within the statistical
errors.
At small $r/\rnod$ one sees deviations from the continuum limit, as one can observe in \fig{alphaqq_artifacts_det}, where only the short distance region is considered. At $r\sim 0.4\rnod$ and $a\sim 0.09\fm$, the deviations can be estimated to about $2\%$ for the Iwasaki action; for the DBW2 action they amount to $4\%$ if one uses $r_{n}$ to define the force, and reach even $40\%$ by employing $\rI$. This fact shows that the adoption of a tree level improved definition of the force does not guarantee success in reducing the lattice artefacts; in particular for RG actions, which are over-corrected at tree level, one should always check the scaling violations for different observables in order to obtain safe statements.

Note that for the DBW2 action the lattice artefacts in $T_{c}\rnod$ are large when one uses $r_{n}$ to define the force. Thus there is not one definition of the force at finite $a$, which has reasonably small lattice artefacts for both $T_{c}\rnod$ and $\alpha_{\qqbar}$.


%% file: glue.tex
\section{Glueball masses}\label{sect_glueball}
For the computation of the glueball masses we followed the method
proposed by \cite{Niedermayer:2000yx}.\\
We decided to concentrate on the states $J^{PC}=0^{++},2^{++}$ by measuring the masses in
the irreducible representations $A_{1}^{++}$, $E^{++}$ and $T_{2}^{++}$ of the cubic group. $A_{1}$ corresponds to the $J=0$ state;
in the continuum limit one expects that the ${\rm O}(3)$ symmetry is restored and hence the doublet $E$ and the triplet $T_{2}$ are degenerate and form together the $J=2$ quintuplet. 

The $\mzpp$ is
particularly interesting for our purpose to investigate the lattice artefacts
on RG actions since these turn out to be quite sizeable for the Wilson
action.\\
We performed a test simulation using the standard Wilson action for
$\beta=6.0$ on a $16^4$ lattice; here we measured all 22 loop shapes up to length 8 that
can be build on the lattice (see \cite{billoire} for the classification of the loops) and formed the wave functions corresponding to the representations 
$R=A_{1}^{++},E^{++},T_{2}^{++}$.
 Our timeslice observable at time $t$ is
\begin{equation}
S_{n}^{R}(t)=\frac{L^{-3/2}}{K}\sum_{{\vec x}}\sum_{i=1}^{d_{n}}c_{n}^{i
  R}\Re W_{n}^{i}({\vec x},t),\quad n=1,...,22,
\end{equation}
where the coefficient $c_{n}^{iR}$ are taken from the literature \cite{billoire,wenger}, the sum over $i$ indicates the
sum over all $d_{n}$ orientations of a given shape $n$ and $K$ is a suitable
normalization constant.\\
Then we build the correlation matrices \footnote{Notice that the vacuum
  subtraction is required only in the $A_{1}^{++}$ channel, since it has the
  same quantum numbers as the vacuum.}
\begin{equation}\label{corrma_glueball}
C_{kl}^{R}(t)=\langle S_{k}^{R}(t)S_{l}^{R}(0)\rangle_{c}=\langle
S_{k}^{R}(t)S_{l}^{R}(0)\rangle-\langle S_{k}^{R}(t)\rangle\langle S_{l}^{R}(0)\rangle.
\end{equation}
The indices $k,l$ assume $(22\times M)$
values, where $M$ is the number of smearing levels. We adopted the same smearing procedure \cite{smear:ape} that we used for the computation of the Wilson loops.

Starting from this large basis, we analyzed the signal/noise ratio of
the different operators in order to eliminate those which introduce large
noise and to reduce the correlation matrix to a set of well
measurable operators.\\
We found general agreement in the classification of the "bad" 
operators with \cite{Niedermayer:2000yx}. Our final choice for the operators to measure was
then
\begin{eqnarray}
A_{1}^{++}  & : & \#2,\#5,\#7,\#8,\#10,\#12,\#14   \\
E^{++}      & : & \#2,\#5,\#7,\#10,\#12,\#14 \\
T_{2}^{++}  & : & \#7,\#12,\#14,
\end{eqnarray}
where for $E^{++}$ and $T_{2}^{++}$ channels we took respectively 2 and 3
orthogonal projections for each shape. For each operator we then considered all
$M$ smearing levels. \\
We assumed that this choice is reasonable for each value of the
lattice spacing and also for the Iwasaki and DBW2 actions and we stress that we did not perform a systematic study on the features of the several operators.\\
The simulation parameters for our new Monte Carlo simulations are listed in
\tab{sim_par_glue}. Measurements were taken after a number of sweeps between 3 and 5.
\begin{table}
\begin{center}
Iwasaki action\\[1ex]
\begin{tabular}{c c c c}
\hline
$\beta$  &  $L$   &  $n_{l}$    &   $n_{meas}$ \\[1ex]    
\hline
2.2423   &  10    &   2,4,6,8   &   12000 \\
2.2879   &  12    &   2,4,6,8   &   16000 \\
2.5206   &  16    &   3,6,9,12  &   8000  \\       
\hline
\hline
\end{tabular}
\\[1ex]
DBW2 action\\[1ex]
\begin{tabular}{c c c c}
\hline
$\beta$  &  $L$   &  $n_{l}$    &   $n_{meas}$ \\[1ex]    
\hline
0.8342   &  12    &   2,4,6,8   &   8000 \\
0.9636   &  16    &   3,6,9,12  &   2500  \\       
\hline
\end{tabular}
\end{center}
\caption{\footnotesize{Simulation parameters for the evaluation of the glueball masses for the Iwasaki and DBW2 actions.}\label{sim_par_glue}}
\end{table}

\subsection{Analysis details}
\begin{figure} 
\begin{center}
\includegraphics[width=7cm]{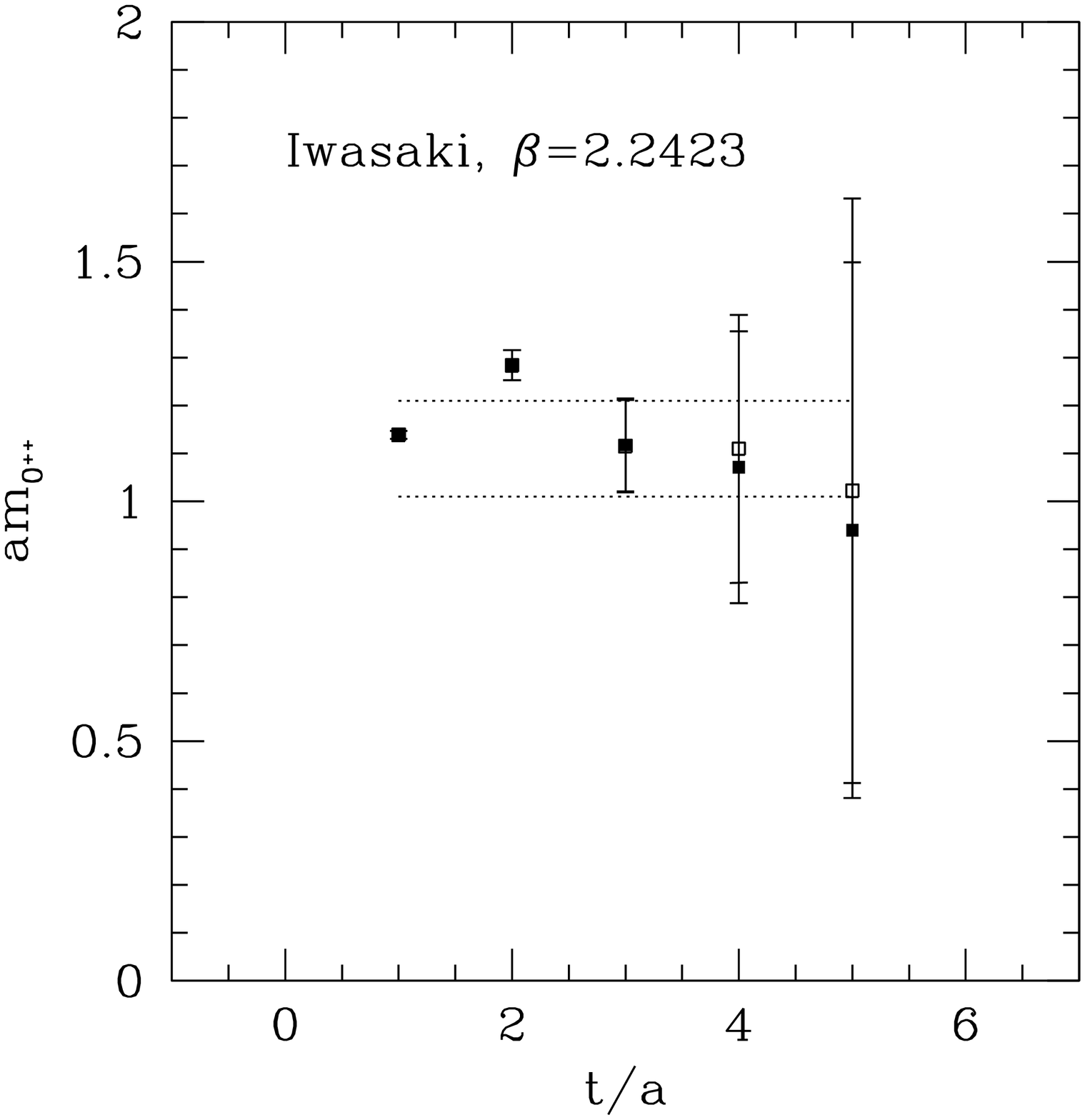} 
\includegraphics[width=7cm]{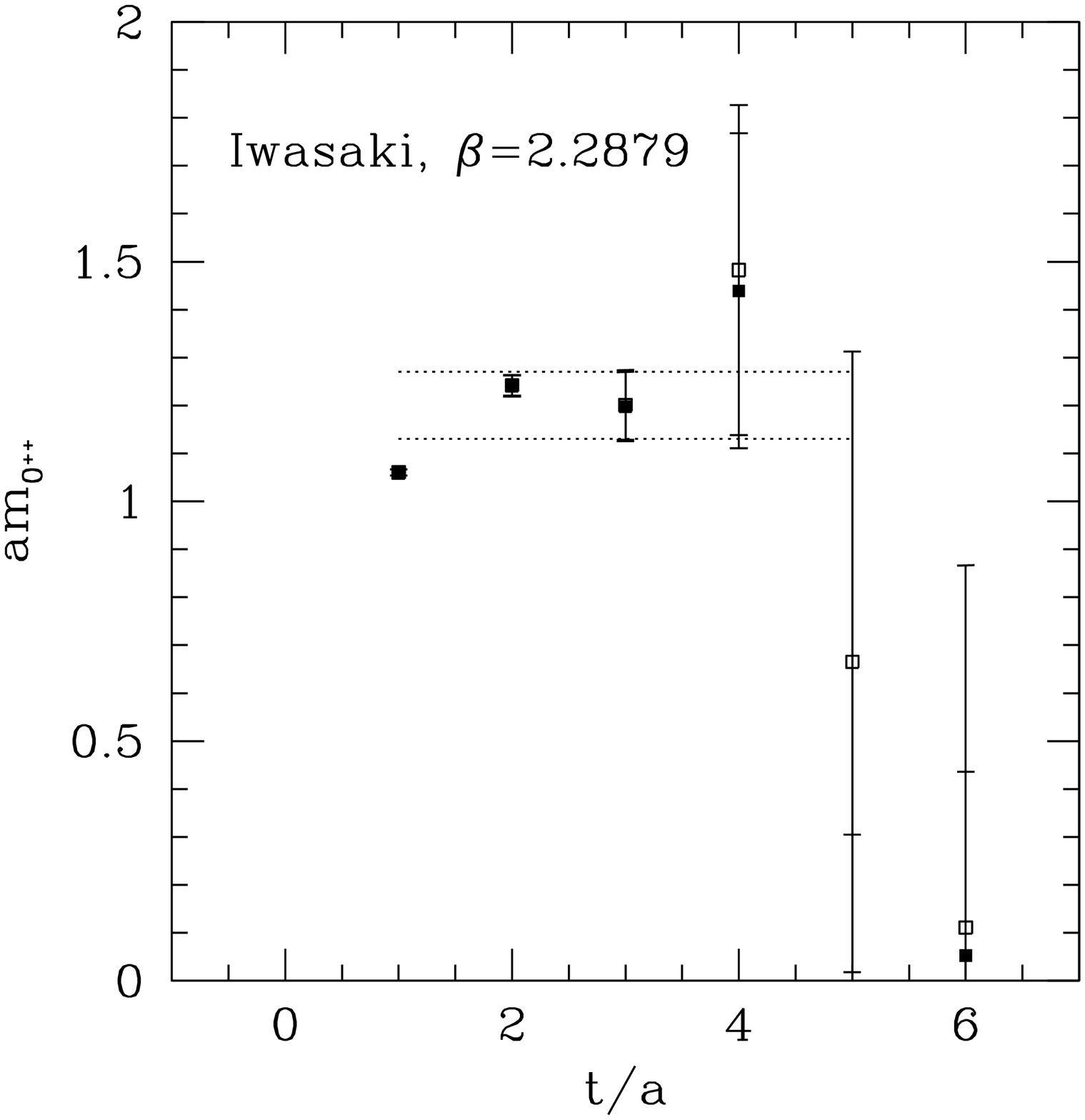} 
\includegraphics[width=7cm]{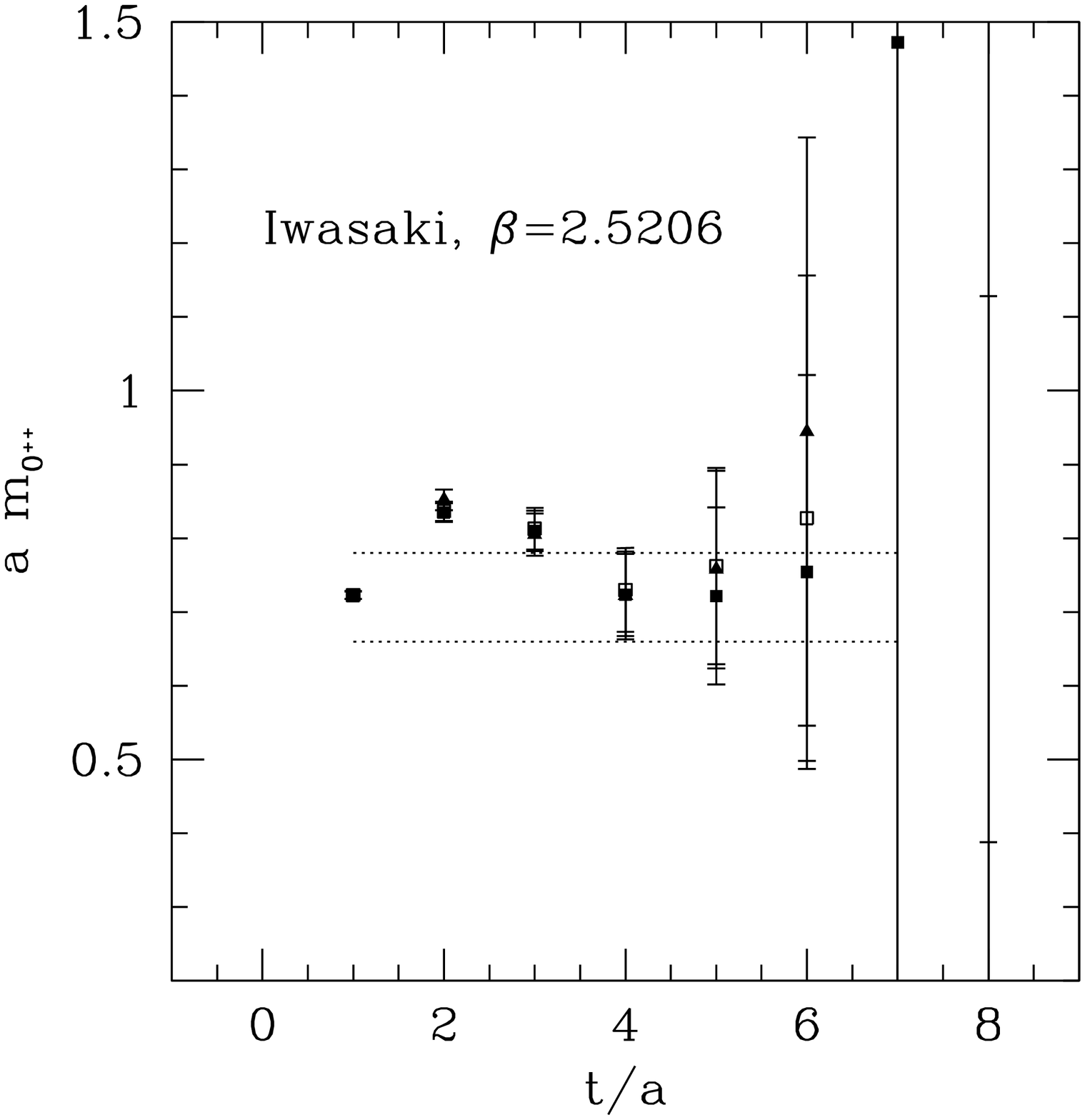} 
\end{center}
\vspace{-0.7cm}
\caption{\footnotesize{The effective masses for the $A_{1}^{++}$ channel,
evaluated with Iwasaki action, at different lattice spacings. The filled and empty squares corresponds to respectively to \protect\eq{meff_eigv} and \protect\eq{meff_corrma} with $t_{0}=0$. For $\beta=2.5206$, the filled triangles correspond to \protect\eq{meff_eigv} with $t_{0}=a$.}\label{fig_meff_iwasaki}}
\end{figure}

\begin{figure}
\begin{center}
\includegraphics[width=7cm]{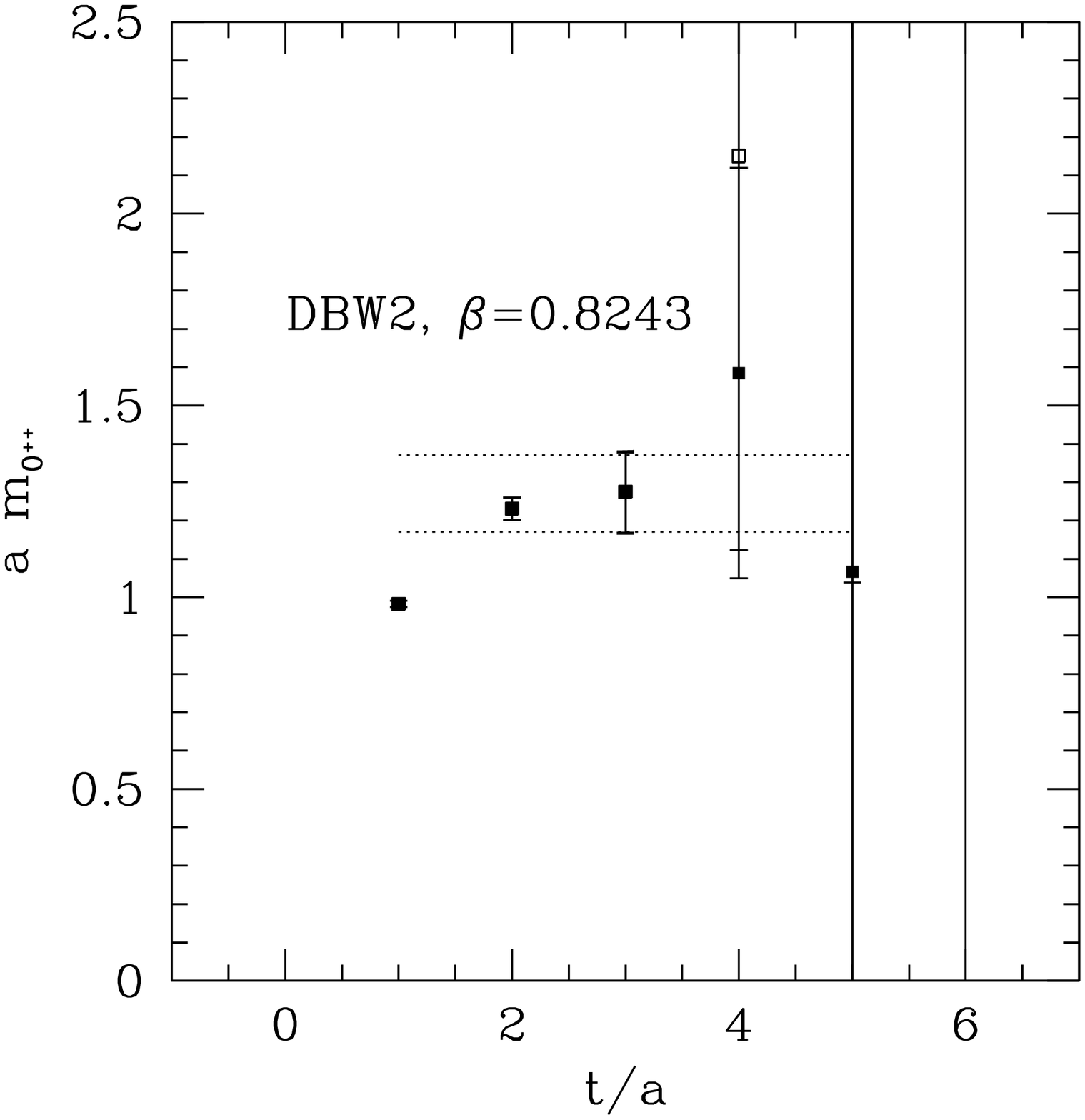} 
\includegraphics[width=7cm]{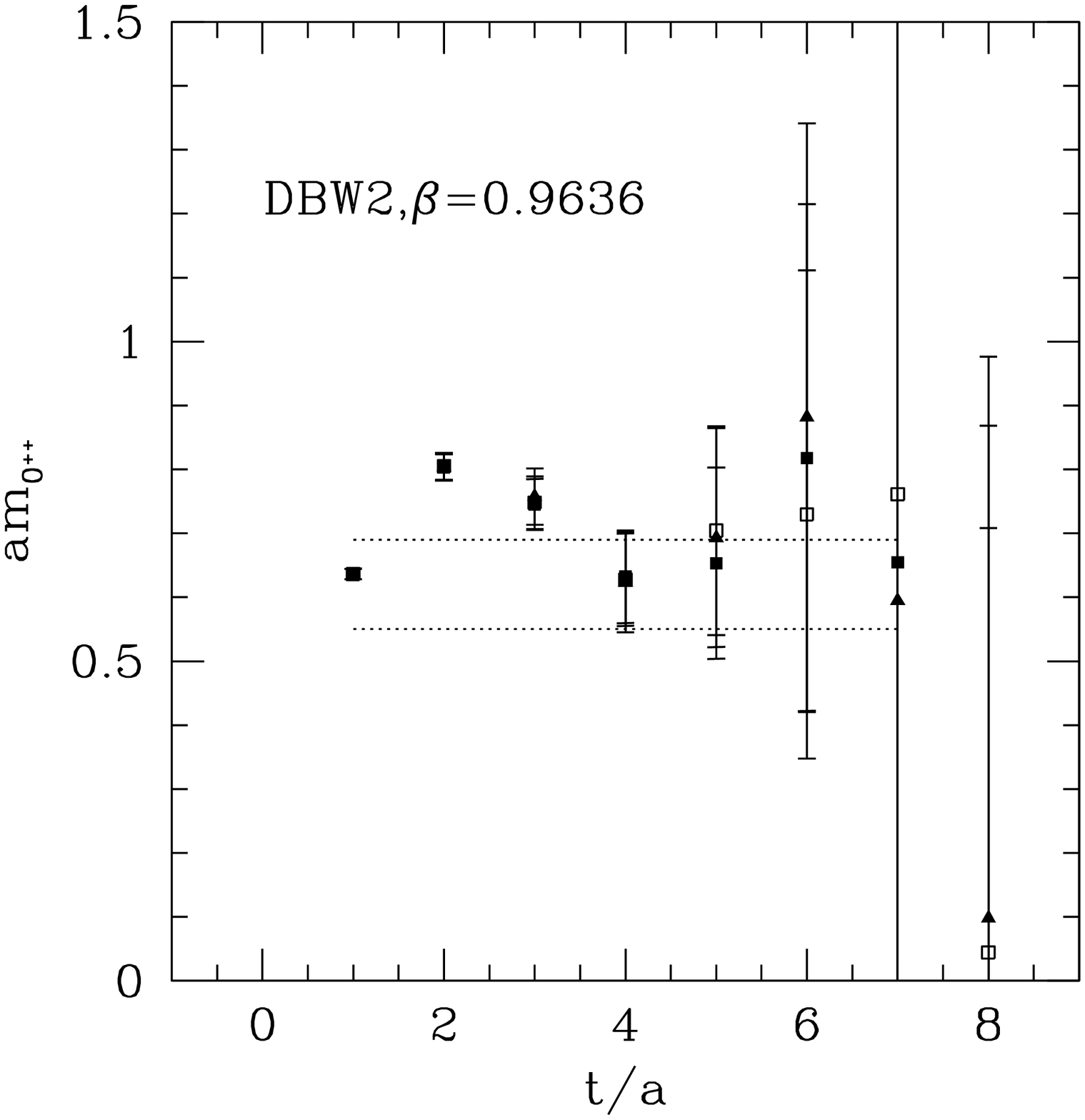} 
\end{center}
\vspace{-0.7cm}
\caption{\footnotesize{The effective masses for the $A_{1}^{++}$ channel,
evaluated with DBW2 action, at different lattice spacings. The filled and empty squares corresponds to respectively to \protect\eq{meff_eigv} and \protect\eq{meff_corrma} with $t_{0}=0$. For $\beta=0.9636$, the filled triangles correspond to \protect\eq{meff_eigv} with $t_{0}=a$.}\label{fig_meff_dbw2}}
\end{figure}

\begin{figure} 
\begin{center}
\includegraphics[width=7cm]{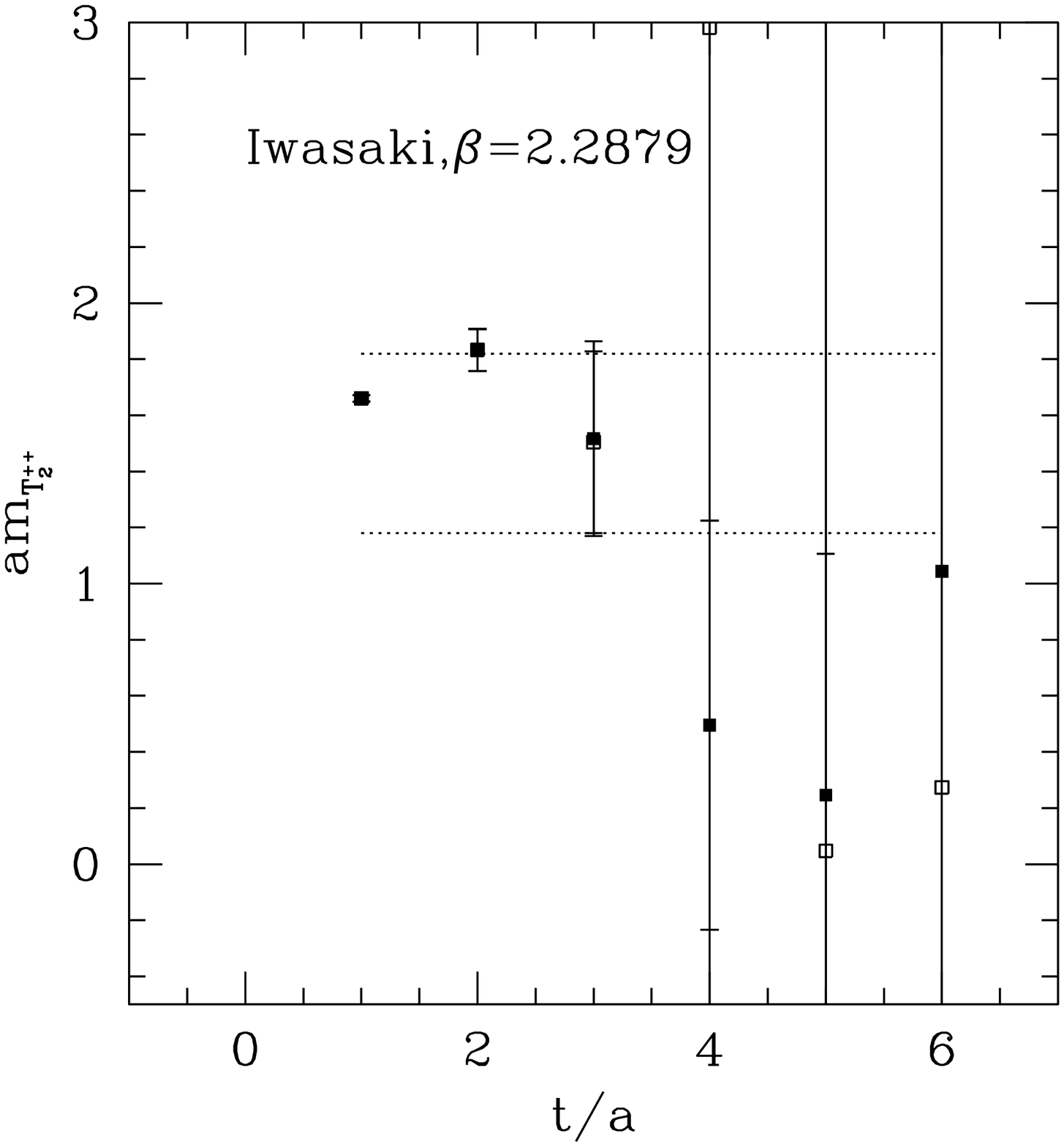} 
\includegraphics[width=7cm]{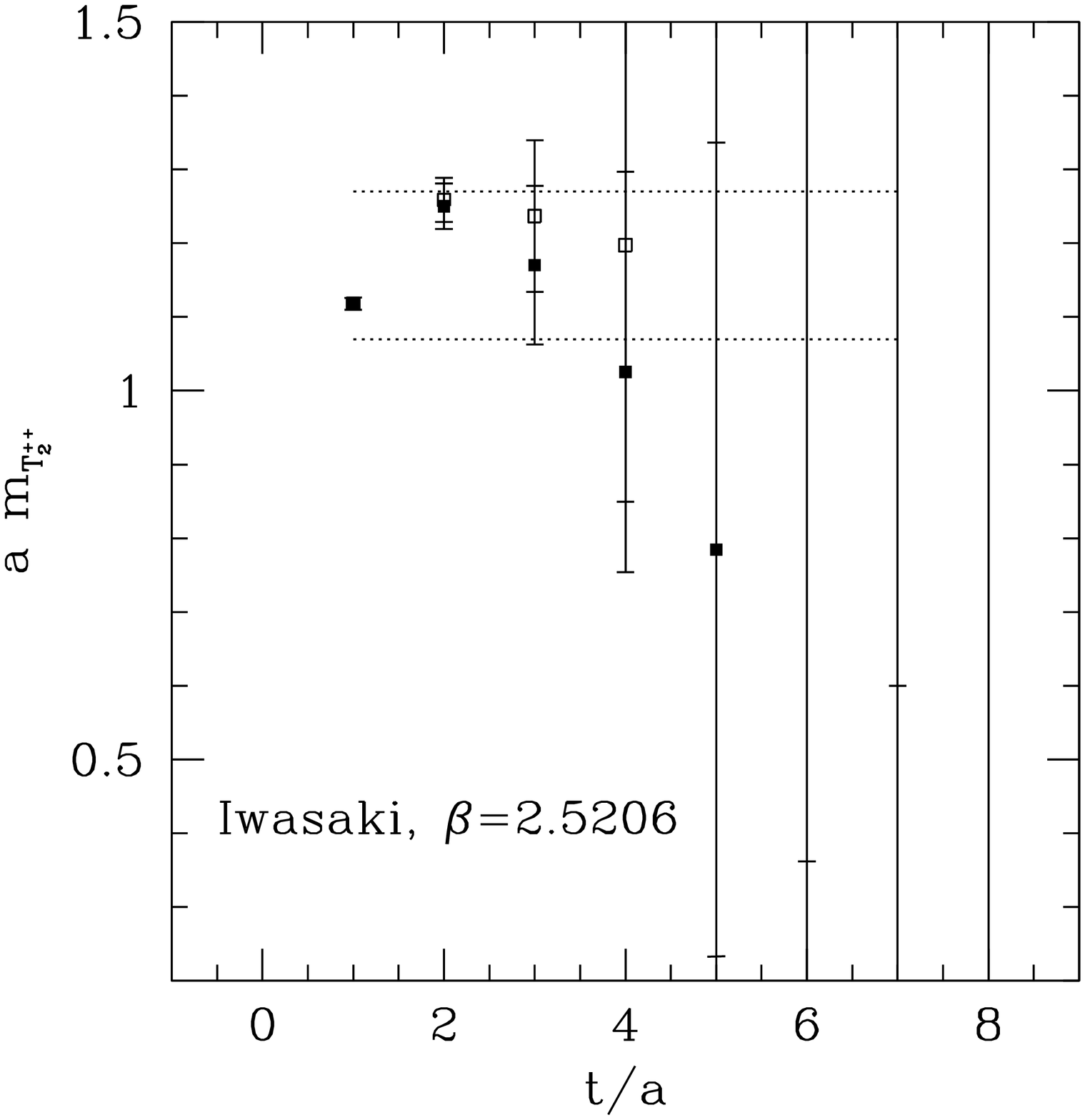} 
\includegraphics[width=7cm]{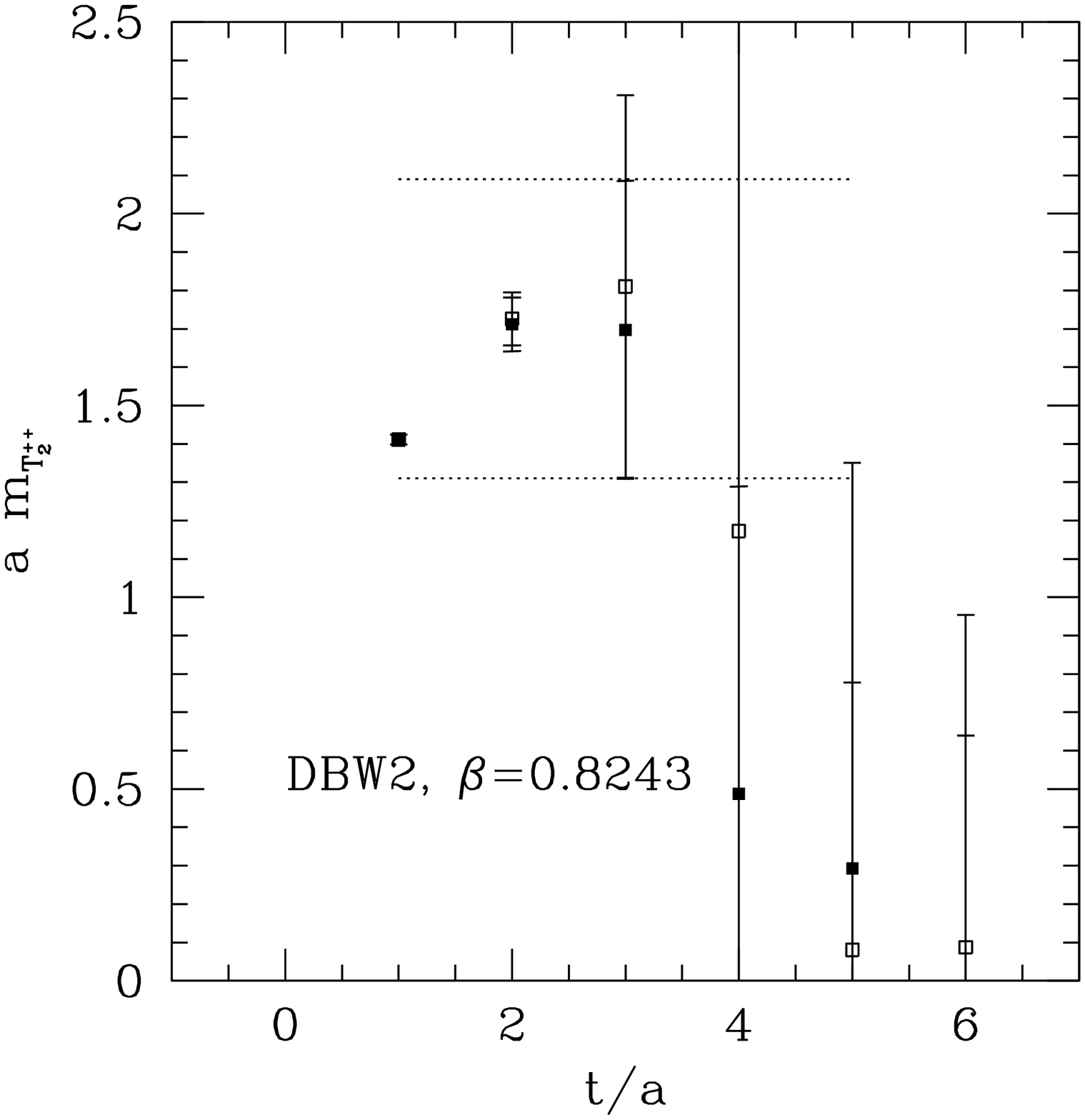} 
\includegraphics[width=7cm]{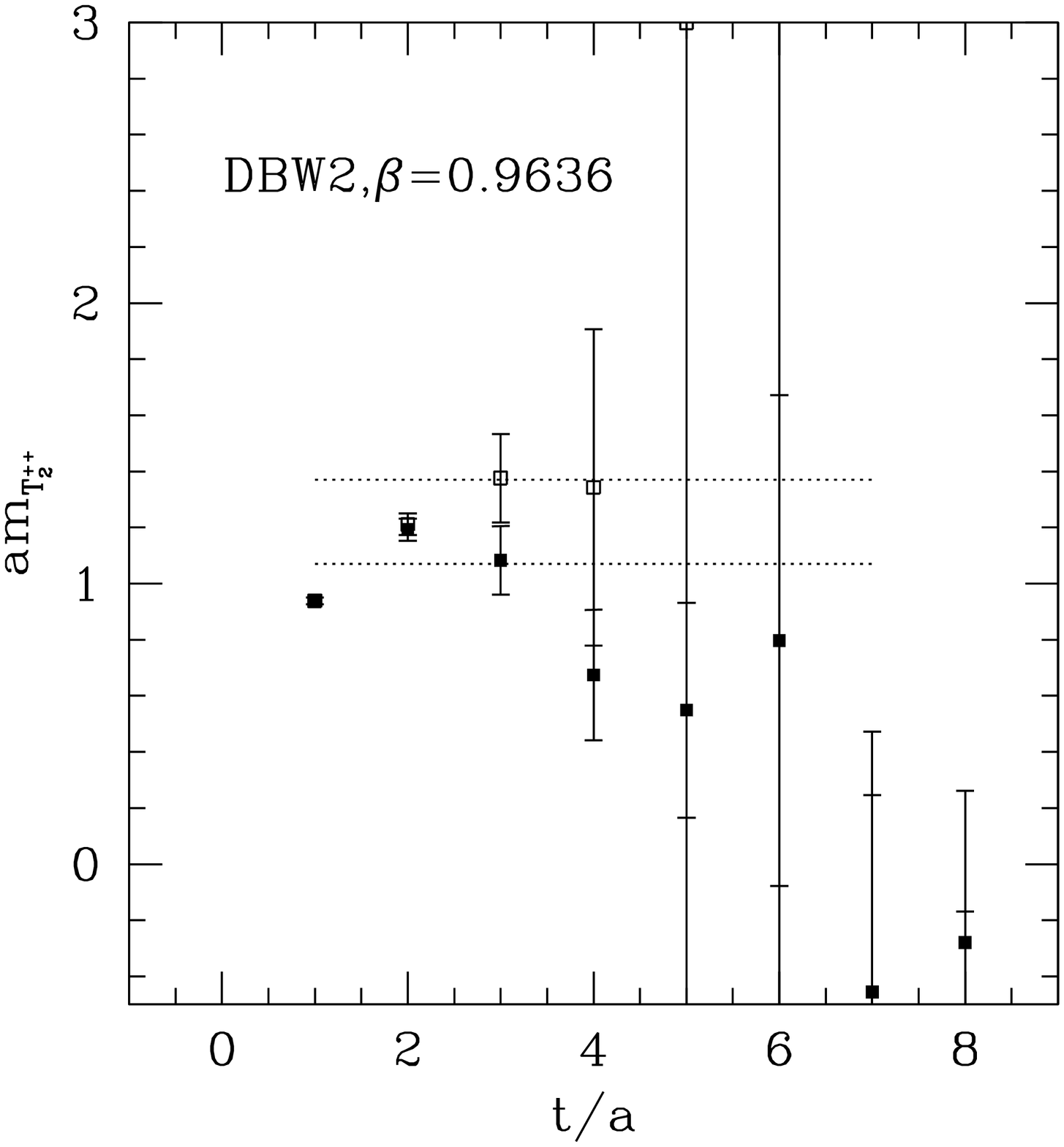} 
\end{center}
\vspace{-0.7cm}
\caption{\footnotesize{The effective masses for the $T_{2}^{++}$ channel,
evaluated with Iwasaki and DBW2 action, at different lattice spacings. The filled and empty squares corresponds to respectively to \protect\eq{meff_eigv} and \protect\eq{meff_corrma} with $t_{0}=0$.}\label{fig_meff2_iwasaki}}
\end{figure}

As already explained in sec. \ref{ppp}, due to the
violation of physical positivity for the RG actions (see app.\ref{app_trans}), the
variational method \cite{varia:michael,phaseshifts:LW} is mathematically not well founded, at least not at small time separations, where one would like to apply it. The statistical errors are indeed
drastically increasing with $t$ and already at $t=4a$ the signal in the
correlation function \eq{corrma_glueball} is lost in noise.\\
We decided to follow \cite{Niedermayer:2000yx} by first solving the generalized eigenvalue problem
\begin{equation}\label{gen_eigv_first}
C(t_{1})v_{\alpha}(t_1,t_0)=\lambda_{\alpha}(t_{1},t_{0})C(t_{0})v_{\alpha}(t_1,t_0),
\end{equation}
with $t_{0}=0$, $t_{1}=a$. Then we projected the correlation matrices to the
space of eigenvectors corresponding to the $N$ eigenvalues which satisfy the
condition \footnote{In \cite{Niedermayer:2000yx} the direct eigenvalues and eigenvectors of $C(t_{0})$ are considered, instead of the generalized one. We have tried both possibilities and found consistent results.}
\begin{equation}\label{epsilon_eigv}
\lambda_{\beta}>\epsilon,\quad\beta=1,...,N,
\end{equation}
where $\epsilon$ is an adjustable (small) parameter. So the reduced matrix is
obtained by
\begin{equation}
C_{ij}^{N}(t)=(v_{i}(t_{1},t_0),C(t)v_{j}(t_{1},t_0)),\quad i,j=1,...,N,
\end{equation}
where the index $R$ for the representation in now omitted.\\
By choosing $\epsilon$ appropriately in \eq{epsilon_eigv} one hopes to get rid
of most of the unphysical modes caused by negative and very small eigenvalues.\\
Then one can apply the variational method to the reduced matrix
\begin{equation}
C^{N}(t)w_{\beta}(t,t_{0})=\lambda_{\beta}(t,t_{0})C^{N}(t_{0})w_{\beta}(t,t_{0}).
\end{equation}
The effective glueball mass can be read off directly from the largest eigenvalue corresponding to the lowest energy
\begin{equation}\label{meff_eigv}
m_{eff}(t)=-\log\left(\frac{\lambda_{0}(t,t_{0})}{\lambda_{0}(t-a,t_{0})}\right).
\end{equation}
or, alternatively one can project again $C^{N}$ to the subspace corresponding to the largest eigenvalue
\begin{equation}
C^{1}(t)=(w_{1},C^{N}(t)w_{1}),
\end{equation}
where $w_{1}=w_{1}(t_{0}+a,t_{0})$;
then one evaluates the glueball masses by
\begin{equation}\label{meff_corrma}
m_{eff}(t)=-\log\left(\frac{C^{1}(t)}{C_{1}(t-a)} \right).
\end{equation}
We tested that these two different evaluation yield results which are
compatible within the statistical errors.\\
We chose $\epsilon$ in \eq{epsilon_eigv} so that the reduced matrix had
dimension between 2 and 5.\\
We applied the same procedure also for $t_{0}=a$, $t_{1}=2a$ in \eq{gen_eigv_first}, but due to the fact that the statistical fluctuations are already quite large one has to start from the beginning from a reduced number of operators in order to be able to solve the generalized eigenvalue problem.\\
We decided to extract the glueball masses by taking the effective mass at
$t=3a$ for the small $\beta$ regime and $t=4a$ for the large value of the
coupling at our disposal. For the RG actions we performed numerical simulations up to $L=16$, with a minimum lattice spacing $a\sim 0.1\fm$, which is quite large and hence we will not be able to perform a continuum extrapolation of the results. 
On the other hand, this is the regime of lattice spacings that have been used for simulations with dynamical fermions and for this reason it is desirable to obtain informations about the discretization errors in this region.\\
The figures \ref{fig_meff_iwasaki},\ref{fig_meff_dbw2} show the effective masses in the $A_{1}^{++}$ channel
computed by using \eq{meff_eigv} (filled squares) and \eq{meff_corrma} (empty squares) for the Iwasaki and the DBW2 action, with
$t_{0}=0$. For the largest value of $\beta$, we show also the results obtained with $t_{0}=a$ (filled triangles).
As for the potential, one notices also in this case the presence
of negative contributions in the correlation functions due to unphysical
states; the situation here is somehow more drastic since one has to discard
for this reason the small $t$ region but on the other hand the errors
increase very rapidly and one is forced to extract the glueball mass at
$t=(3-4)a$. The results can then be affected by systematic errors that
can not be easily estimated. In the plots we have indicated 
with the dotted lines the range in which we decided to take the mass.\
The statistical errors were evaluated by using a jackknife procedure.\\
Concerning the determination of $m_{2^{++}}$ we observed that the signal for the $E^{++}$ channel is usually worse than for the $T_{2}^{++}$ and the errors on the effective masses are very large already at $t=3a$. For this reason we decided to use  $m_{T_{2}^{++}}$ as estimate of $m_{2^{++}}$ at finite lattice spacing.\\
\Fig{fig_meff2_iwasaki} shows the effective masses for the $T_{2}^{++}$ channel. The determination of the masses is more problematic that the $A_{1}^{++}$ case; in particular for the largest coupling one can observe that at $t=4a$ the errors are too large to have a significant measurement of the mass. We decided hence to extract the mass at $t=3a$, taking care that this value is compatible with the one at $t=4a$ within the statistical errors.

\subsection{Results}
\begin{figure}
\begin{center}
\includegraphics[width=11cm]{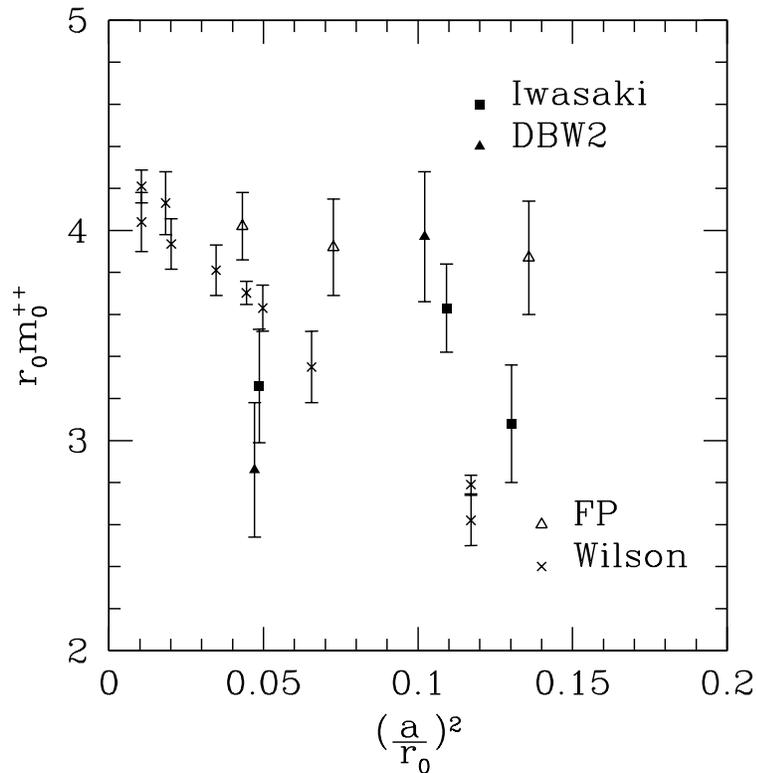} 
\end{center}
\vspace{-1cm}
\caption{\footnotesize{The $0^{++}$ glueball mass normalized with $\rnod$ as
    function of $(a/\rnod)^2$ for different actions.}\label{f_0++}}
\end{figure}

\begin{figure}
\begin{center}
\includegraphics[width=11cm]{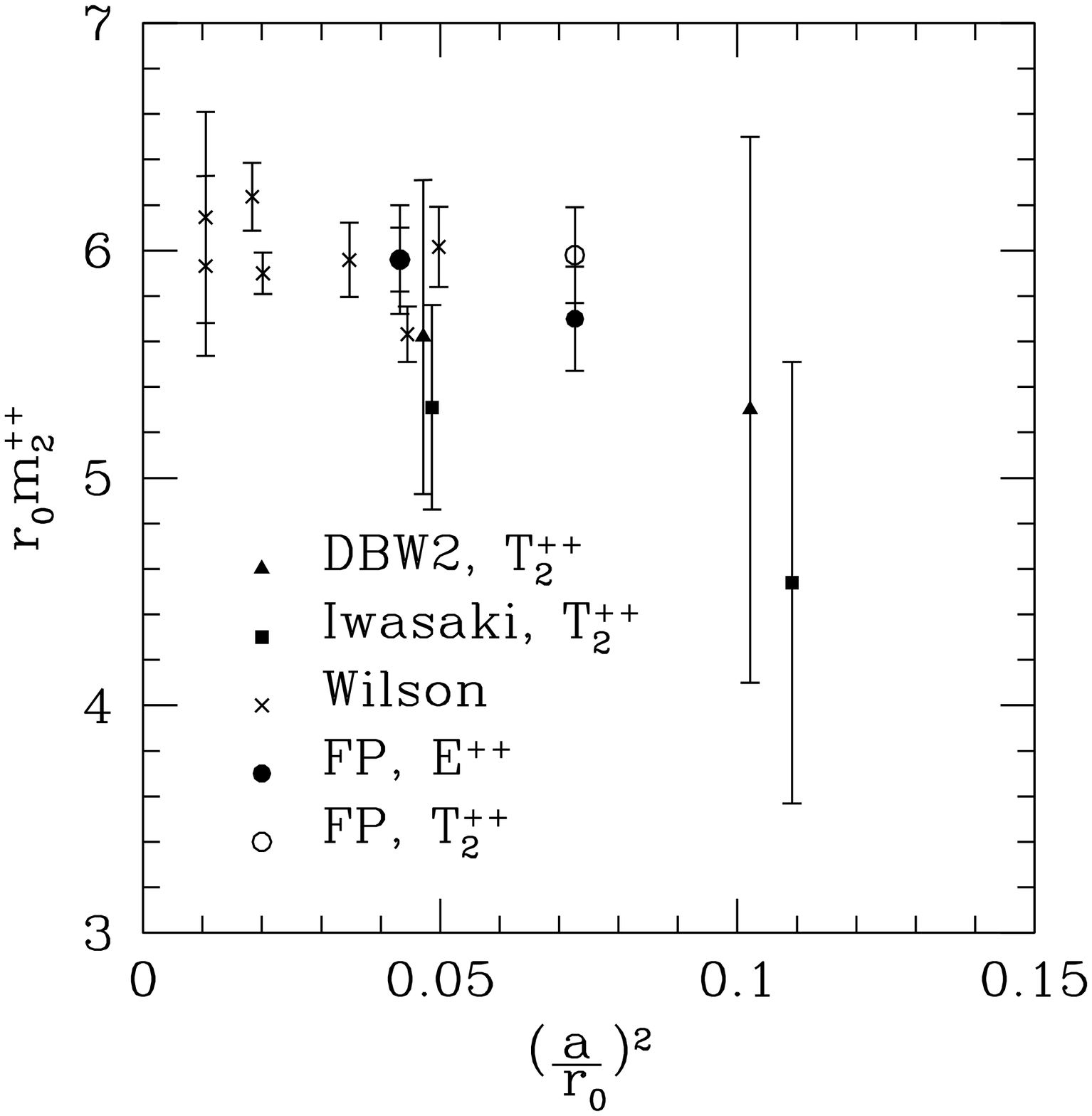} 
\end{center}
\vspace{-1cm}
\caption{\footnotesize{The $2^{++}$ glueball mass normalized with $\rnod$ as
    function of $(a/\rnod)^2$ for different actions.}\label{f_2++}}
\end{figure}

In \tab{glue_res_iwasaki1} the results for the masses in
lattice units are reported. For the smallest $\beta$ for the Iwasaki action,
it was not possible to obtain a reliable evaluation of the mass in the $T_{2}^{++}$ channel.\\
The renormalized quantities $\rnod\m_{G}$ are reported in tables
\ref{glue_res_iwasaki2} and \ref{glue_res_dbw22}; due to the fact that the errors
on $a m_{G}$ are quite large (between $8\%$ and $9\%$ for $A_{1}^{++}$ and between $10\%$ and $20\%$ for $T_{2}^{++}$), the difference which
arises by choosing $r_{I}$ or $r_{n}$ is in this case much smaller than the total
uncertainty; we have reported however both results in the table.
For $\beta=2.2423$ we had no direct measurement of $\rnod/a$ and we made use of the interpolation formula \eq{fit_iwasaki}.\\
\Fig{f_0++} shows the results for $\rnod m_{0^{++}}$ as function of
$(a/\rnod)^2$, displaying only the results obtained with $r_{n}$.
 For the comparison we included the results for FP action \cite{Niedermayer:2000yx}
and several calculations performed with the Wilson action \cite{Ishikawa:1983xg,Vaccarino:1999ku,Bali:1993fb}.\\
The continuum values avaliable in the literature are listed in \tab{glue_cont}
and have been taken from \cite{Niedermayer:2000yx} for the FP action and from
\cite{Wittig:1999kb} for the Wilson action, where the results of \cite{Ishikawa:1983xg,Vaccarino:1999ku,Bali:1993fb} have been
expressed in units of $\rnod$.\\
The interpretation of our results is not very clear, also due to the large errors.\\
In any case, we want to stress that our determination can be seen at least as an upper limit for $m_{0^{++}}$ and $m_{2^{++}}$. We expect that at the values of $t/a$ at which we extracted the masses, the effects of positivity violations 
have already disappeared; this assumption is justified by our study based on perturbation theory and our estimation of $t_{min}$ \eq{tmin}.\\
For this reason we believe that possible systematic uncertainties on the glueball masses could only be due to the presence of excited states and hence could affect our measurement only in such a way that the real values of $m_{0^{++}}$, $m_{2^{++}}$ are lower with respect to our determination.\\
\noindent
At lattice spacings $a\sim 0.15\,\fm$ we notice a
improvement of the RG actions with respect to the Wilson action;
comparing with the continuum limit we find no significant discrepancy both for DBW2 and Iwasaki action,
while for the Wilson action one finds  $30-40\%$ deviation.\\
At lattice spacing $a\sim 0.1\,\fm$ we find on the other hand large lattice artefacts for RG actions: the result obtained with the Iwasaki
action is compatible with the one calculated through the plaquette action at the same lattice spacing,
while for the DBW2 action it is even further away from the continuum limit.\\
If one considers our measurement as upper limit, one could conclude that the RG improved actions are not able to cure the problem of large lattice artefacts for the $0^{++}$ glueball mass. 
At very small lattice spacings one
expects that the dominant lattice artefacts are of order $a^2$; for the
Wilson action this is indeed well confirmed by the numerical results in the
range $a\lesssim 0.17\fm$, as one can see in \fig{f_0++}.\\
For alternative actions there is no reason a priori to observe the same
behavior: while at small lattice spacings the ${\rm O}(a^2)$ should in any
case be the dominant one, at larger $a$ it is possible that lattice artefacts
are governed by higher orders (${\rm O}(a^4)$ and higher).\\
We have indeed already observed deviations from ${\rm O}(a^2)$ behavior in RG
actions for the quantity $T_{c}\rnod$ and even already for the force computed
at tree level in the previous sections.\\
%
In \fig{f_2++} we report our results for $\rnod m_{2^{++}}$; 
for this particular observable the calculation performed with the Wilson action do not show significant lattice artefacts.\\
At our smallest lattice spacing we do not observe a deviation from the results obtained with the Wilson action; one has however to notice that our errors are too large to make any conclusive statement.\\
For our largest lattice spacing the results with the Wilson action are not available (there are only know results with anisotropic lattices \cite{Morningstar:1999rf}) and it is indeed difficult to have a reliable estimation of the mass also in our case: we decided to show however our results, even with the very large error bars.\\
From our computation one can not deduce if for $\rnod m_{2^{++}}$ the RG actions show significative discretization errors  
 and further investigations are needed to clarify the issue.\\
In particular, the exponential error reduction proposed by L\"uscher and Weisz in \cite{Luscher:2001up} has been already tested for the evaluation of the $0^{++}$ and $2^{++}$ masses in \cite{Meyer:2002cd} yielding promising results. We expect that the implementation of this algorithm for RG actions could help in reducing the errors for the correlation functions at quite large $t$ and hence to have more reliable estimations of the glueball masses.\\


\begin{table}
\begin{center}
\begin{tabular}{c c c}
\hline
Collab.  & $\rnod m_{0^{++}}$   &  $\rnod m_{2^{++}}$ \\[1ex]
\hline
M \& P \cite{Morningstar:1999rf}   & 4.21(11)(4)  & 5.85(2)(6) \\   
GF11 \cite{Vaccarino:1999ku}     & 4.33(10)     & 6.04(18)    \\
Teper\cite{glueb:teper98}     & 4.35(11)     & 6.18(21)     \\
UKQCD\cite{Bali:1993fb}     & 4.05(16)     & 5.84(18)   \\
FP  \cite{Niedermayer:2000yx}      & 4.12(21)     & [5.96(24)]      \\[1ex]
\hline
\end{tabular}
\end{center}
\caption{\footnotesize{Continuum extrapolations of the two lowest glueball
    masses in units of $\rnod$. For the FP action, the $2^{++}$ value is not extrapolated to the continuum but denotes the mass obtained at a lattice spacing $a=0.10\,\
fm$.}\label{glue_cont}}
\end{table}
\begin{table}
\begin{center}
Iwasaki action\\[1ex]
\begin{tabular}{c c c}
\hline
$\beta$  & $am_{A_{1}^{++}}$ & $am_{T_{2}^{++}}$ \\[1ex]
\hline
2.2423   & 1.11(10)          &                             \\
2.2879   & 1.20(7)     &    1.50(32) \\
2.5206   & 0.72(6)      &   1.17(10)      \\   
\hline
\end{tabular}
\end{center}
\begin{center}
DBW2 action\\[1ex]
\begin{tabular}{c c c}
\hline
$\beta$  & $am_{A_{1}^{++}}$ &   $am_{T_{2}^{++}}$ \\[1ex]
\hline
0.8243   & 1.27(10)               &  1.70(39)  \\
0.9636   & 0.62(7)                &  1.22(15)   \\
\hline
\end{tabular}
\end{center}
\caption{\footnotesize{Glueball masses in lattice units, Iwasaki and DBW2 action.}\label{glue_res_iwasaki1}}
\end{table}
\begin{table}
\begin{center}
Iwasaki action\\[1ex]
\begin{tabular}{c c c c c}
\hline
$\beta$  & $\rnod m_{A_{1}^{++}}$ ($r_n$) &  $\rnod m_{T_{2}^{++}}$ ($r_n$) & $\rnod m_{A_{1}^{++}}$ ($\rI$) &  $\rnod m_{T_{2}^{++}}$ ($\rI$) \\[1ex]
\hline
2.2423   &  3.08(28)  &       &  3.07(28)  &                           \\
2.2879   &  3.63(21)        & 4.54(97)  &  3.63(21)       &   4.54(97)   \\
2.5206   &  3.26(27)        & 5.31(45)  &  3.25(27)       &   5.28(45)   \\[1ex]
\hline
\end{tabular}
\end{center}
\caption{\footnotesize{Results for $\rnod m_{G}$ for the channels $A_{1}^{++}$
    and $T_{2}^{++}$, using the Iwasaki action. }\label{glue_res_iwasaki2}}
\end{table}
\begin{table}
\begin{center}
DBW2 action\\[1ex]
\begin{tabular}{c c c c c}
\hline
$\beta$  & $\rnod m_{A_{1}^{++}}$ ($r_n$) &  $\rnod m_{T_{2}^{++}}$ ($r_n$) & $\rnod m_{A_{1}^{++}}$ ($\rI$) &  $\rnod m_{T_{2}^{++}}$ ($\rI$) \\[1ex]
\hline
0.8243   &  3.97(31)         &   5.3(1.2)    &  3.86(30)         &  5.2(1.2)              \\
0.9636   &  2.86(32)         &   5.62(69)    &  2.82(32)         &  5.56(68)              \\[1ex]
\hline
\end{tabular}
\end{center}
\caption{\footnotesize{Results for $\rnod m_{G}$ for the channels $A_{1}^{++}$
    and $T_{2}^{++}$, using the DBW2 action.}\label{glue_res_dbw22}}
\end{table}


%% file: disc.tex
\section{Discussion}
The adoption of alternative gauge actions with the purpose to reduce the lattice artefacts and/or to improve chiral properties in QCD simulations must be accompanied with an accurate study of their properties. In particular, universality and scaling behavior must be tested on a large set of observables.\\
We have chosen $\rnod$, the deconfining temperature and the glueball masses
$m_{0^{++}}$, $m_{2^{++}}$ to perform this test for the Iwasaki and DBW2
actions, which contain the same operators  with different coefficients.\\
We computed the scale $\rnod$ through the force between static quarks at several values of $\beta$ corresponding to
the critical couplings for deconfinement and finally built the renormalized
quantity $T_{c}\rnod$.  By approaching the continuum limit we verified the
universality between Wilson and RG action, which had been doubted in previous
calculations where the string tension $\sigma$ was used to set the scale
instead of $\rnod$. Our new computation confirms that $\rnod$ is a more
appropriate quantity, since the systematic uncertainties are reduced with
respect to the string tension.\\
From the point of view of the scaling behavior, for this particular quantity
the Iwasaki action is able to reduce the lattice artefacts. For the DBW2 action, on the other hand,
we observe large lattice artefacts at $a\sim .....$. While they can be made small in $\rnod T_{c}$ 
by choosing a $\rI$ in the definition of the force,this choice leeds to large $a$-effects in $\alpha_{\qqbar}$. The coefficient $c_{1}$ appears to be too far away from its tree-level value.
By investigating the force at tree level we pointed out that RG actions are
``over-corrected'', in the sense that they can introduce even larger lattice
artefacts than the Wilson action. This suggests that the improvement may work
only in specific coupling ranges; moreover, it can happen that the ${\rm
  O}(a^2)$ scaling violation is dominant only at very small lattice spacings,
and this feature requests a particular care in performing
the continuum extrapolation.\\
Another important fact that we stressed is the violation of physical
positivity which occur in RG actions, and more generally in actions that
contain other terms apart from the plaquette. We clearly observed the presence
of unphysical states in the evaluation of the static quark potential from
Wilson loop correlation functions.\\
The main problem connected with this fact is the failure in the mathematical
assumptions which allow to apply the variational method.
The method can be safely applied only if one manages to get rid of the
unphysical modes by going at sufficiently large time separations - but here
the statistical errors increase dramatically - or by finding a reliable
procedure to project them out from the correlation matrix.\\
We followed \cite{Luscher:1984is} and estimated in perturbation theory at which time
separation these lattice artefacts are expected to disappear and this
indicative study is confirmed by our numerical evaluation of the static
potential.\\
We stress that we did not perform a systematic study on the efficiency of the
smearing procedure for RG actions and we applied the same criteria adopted for
the Wilson action at analogous lattice spacings. The presence of states with
negative norm can spoil the search for an ``optimal'' smearing since it can
mask the presence of excited states at small time separations, and a
reasonable criterium must be found.\\
In a second part of the work we performed numerical simulations to evaluate $m_{0^{++}}$
and $m_{2^{++}}$ at several lattice spacings for RG actions. The main
motivation was that the $0^{++}$ glueball show large scaling deviations and
hence is an ideal observatory to make a comparison with alternative actions.\\
Here the violation of positivity constitutes an even stronger problem, since for these observables the signal is lost in noise already at small time separations.\\
Due to these difficulties, our estimations of glueball masses are affected by large errors and hence not suitable to draw a definitive conclusion about the lattice artefacts; a possible solution would be to apply specific algorithms for the variance reduction in order to extend the precision on the effective masses to larger values of $t$, where the unphysical states are supposed to be absent.\\
The fact that for the Wilson action the lattice artefacts for $m_{0^{++}}$ are large and the glueball mass in lattice units becomes small is usually interpreted as the ``influence'' of the endpoint of the first order phase transition which for $\SUthree$ plaquette action is located in the fundamental-adjoint coupling plane at $(\beta_{f},\beta_{a})=(4.00(7),2.06(8))$ \cite{Blum:1995xb}.\\
This issue turns out to be very important for the next simulation with dynamical fermions; in particular, in $O(a)$ improved Wilson fermions it is believed that the clover term produces an adjoint term in the gauge action with positive coupling which could be responsible for the observed first order phase transition at zero-temperature in the $N_{f}=3$ simulations \cite{Aoki:2001xq}. 
The JLQCD investigations indicate that this
transition is a lattice artefact restricted to strong coupling
regions ($\beta\lesssim 5.0$) and it disappears if one adopts
other gauge actions, like Iwasaki or tadpole-improved Symanzik action, or alternatively if one uses the unimproved Wilson fermionic action
with the plaquette gauge action.\\
By considering our estimation as an upper limit for the glueball masses, it seems that RG actions do not lead to a sensible reduction of the lattice artefacts for $m_{0^{++}}$, and hence the scenario remains unclear. 
A possible alternative to avoid the problem would be to adopt the usual plaquette action with a negative adjoint term, and studies on the scaling behavior for this kind of actions are ongoing.


%% file: app_trans.tex
\section{Positivity violation in improved gauge actions}\label{app_trans}
In \cite{Luscher:1984is} it has been pointed out that for lattice gauge actions containing general classes of loops it is still possible to construct a transfer matrix $\mathcal{T}$. 
In particular, restricting to actions of kind \eq{impr_action} with plaquettes and planar $(1 \times 2)$ loops, the transfer matrix can be defined as transition amplitude between fields configurations defined on two consecutive pairs of equal time hyperplanes.
But, unlike the Wilson case, one finds that $\mathcal{T}$ is no longer hermitean, and its spectral values $\lambda$ occur in pairs of complex conjugated numbers.\\
Moreover, one can show that the spectrum $\sigma(\mathcal{T})$ contains a real, positive, non-degenerate eigenvalue $\lambda_{0}$ such that\\
\begin{equation}
\lambda_{0}>|\lambda|,\quad\forall\lambda\in\sigma(\trans),
\quad\lambda\neq\lambda_{0}.
\end{equation}
The corresponding eigenfunction may be interpreted as ground state wave function.\\
The violation of physical positivity forbids the definition of an Hamiltonian operator; from the point of view of the spectral decomposition of connected two-point functions, one still observes expontential decay at large time-separations, but contributions with negative weight will appear, as remnant of the positivity violation.\\
Nevertheless, the violation of positivity is a lattice artefact and one expects that physical positivity is recovered in the continuum limit; one should be able to define a subspace of states with small energy with respect to the cut-off where the non-hermiticity of the trasfer matrix can be eliminated.\\
This corresponds to the fact that for large enough time-separations, the negative contributions to the spectral decomposition of two-point functions should disappear.\\ 
Following \cite{Luscher:1984is}, one can argue that there exists a $0<\epsilon<1$  such that independently of the cutoff the following properties hold:\\
\noindent
(i) all spectral values $\lambda\in\sigma(\trans)$ with $|\lambda|\geq\epsilon\lambda_{0}$ are real and positive; \\
\noindent
(ii) one can define a physical Hilbert space $\mathcal{H}_{phys}\in\mathcal{H}$ and a new scalar product such that
$(\Psi,\Psi)_{new}>0,\quad\forall\Psi\in\mathcal{H}_{phys},\quad\Psi\neq 0$.\\
The existence of such $\epsilon$ has not been rigorously proved, but it is
supported for example by perturbative calculations.
In the next section we will estimate $\epsilon$ by determining the location of unphysical poles in the free propagator.

\subsection{Unphysical poles in the propagator}
\begin{figure}
\begin{center}
\includegraphics[width=9cm]{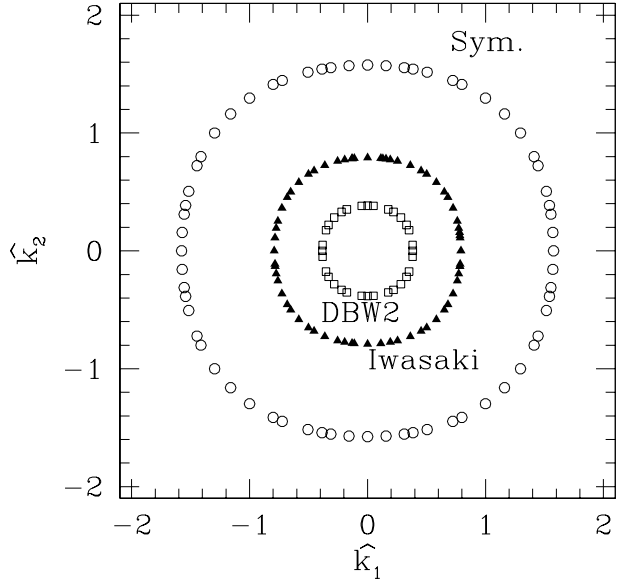}
\end{center}
\vspace{-1cm}
\caption[The curves defined by the condition $\mathfrak{Im}(w)=0$ in the $\hat{k}_{3}=0$ plane. Inside the curve $\mathfrak{Im}(w)=0$, while outside $\mathfrak{Im}(w)\neq 0$.]{\footnotesize{\label{brill}The curves defined by the condition $\mathfrak{Im}(w)=0$ in the $\hat{k}_{3}=0$ plane. Inside the curve $\mathfrak{Im}(w)=0$, while outside $\mathfrak{Im}(w)\neq 0$.}}
\end{figure}

In \cite{Weisz:1983zw} the propagator $D_{\mu\nu}$ associated to the action \eq{impr_action} is evaluated to lowest order in $g_{0}$ with covariant gauge fixing; it is defined by
\begin{equation}\label{pert_propagator}
\langle A_{\mu}^{i}(x) A_{\nu}^{j}(y)\rangle=\delta^{ij}\int_{k}e^{ik(x-y)}e^{i(k_{\mu}-k_{\nu})/2}D_{\mu\nu}(k),
\end{equation}
where $A_{\mu}(x)$ is related to the link variables by 
$$
U(x,\mu)=e^{-A_{\mu}(x)}
$$
and the momenta integration corresponds to
$$
\int_{k}=\prod_{\mu=0}^{3}\int_{-\pi}^{\pi}\frac{dk_{\mu}}{2\pi}.
$$ 
We are now referring to the case $L=\infty$, so that the momenta take continuous values in the Brillouin zone.\\
\Eq{pert_propagator} can be rewritten in the form 
\begin{equation}\label{pert_propagator2}
D_{\nu\tau}(k)=(\hat{k}^{2})^{-2}\left[\alpha\hat{k}_{\nu}\hat{k}_{\tau}+\sum_{\sigma}(\hat{k}_{\sigma}\delta_{\tau\nu}-\hat{k}_{\tau}\delta_{\sigma\nu})A_{\tau\sigma}(k)\hat{k}_{\sigma}\right],
\end{equation}
where $\alpha$ is the gauge parameter, $\hat{k}_{\mu}=2\sin(k_{\mu}/2)$ and 
$A_{\mu\nu}$  is independent of $\alpha$ and has the general form
\begin{equation}
A_{\mu\nu}=\frac{f(\hat{k})}{D},
\end{equation}
with
\begin{equation}
D=\sum_{\mu}\hat{k}_{\mu}^{4}\prod_{\nu\neq\mu}q_{\mu\nu}+\sum_{\mu>\nu,\rho >\tau,\{\rho,\tau\}\cup\{\mu,\nu\}=\emptyset}\hat{k}_{\mu}^{2}\hat{k}_{\nu}^{2}q_{\mu\nu}(q_{\mu\rho}q_{\nu\tau}+q_{\mu\tau}q_{\nu\rho}).
\end{equation} 
The explicit form of $f(\hat{k})$ can be found in \cite{Weisz:1983zw} and will not reported here.
The denominator of \eq{pert_propagator2} is then given by
\begin{equation}\label{delta}
\Delta=D(\hat{k}^2)^2=
\end{equation}
$$
(\hat{k}^2)^2\left(\sum_{\mu}\hat{k}_{\mu}^{4}\prod_{\nu\neq\mu}q_{\mu\nu}+\sum_{\mu>\nu,\rho >\tau,\{\rho,\tau\}\cup\{\mu,\nu\}=\emptyset}\hat{k}_{\mu}^{2}\hat{k}_{\nu}^{2}q_{\mu\nu}(q_{\mu\rho}q_{\nu\tau}+q_{\mu\tau}q_{\nu\rho})\right),
$$
where $q_{\mu\nu}$ for our specific form of the action \eq{impr_action} takes the form
\begin{equation}
q_{\mu\nu}=(1-\delta_{\mu\nu})[1-c_{1}(\hat{k}_{\mu}^{2}+\hat{k}_{\nu}^{2})].
\end{equation}
Then \eq{delta} becomes
\begin{equation}
\Delta=\left(\hat{k}^{2}-c_{1}\sum_{\mu}\hat{k}_{\mu}^{4}\right)\Bigg[\hat{k}^{2}-c_{1}\left((\hat{k}^{2})^{2}+\sum_{\mu}\hat{k}_{\mu}^{4}\right)+
\end{equation}
$$
\frac{1}{2}c_{1}^{2}\left((\hat{k}^{2})^{3}+2\sum_{\mu}\hat{k}_{\mu}^{6}-\hat{k}^{2}\sum_{\mu}\hat{k}_{\mu}^{4}\right)\Bigg]
-4c_{1}^{3}\sum_{\mu}\hat{k}_{\mu}^{4}\prod_{\nu\neq\mu}\hat{k}_{\nu}^{2}.
$$
In order to search for the poles of the propagator,
one substitutes\\
$$
k=(\hat{k}_{1},\hat{k}_{2},\hat{k}_{3},iw)                                     $$
and looks for solutions of the equation $\Delta=0$ scanning the whole Brillouin zone. In general for $c_{1}\neq 0$ one expects complex conjugated solutions
$$                                       
w=\mathfrak{Re}(w) \pm i\mathfrak{Im}(w)=w(\hat{k}_{1},\hat{k}_{2},\hat{k}_{3}).
$$
Numerical investigations showed that the condition
$\mathfrak{Im}(w)=0$ (for \emph{all} solutions at a given momentum) defines 
the equation $f(\hat{k}_{1},\hat{k}_{2},\hat{k}_{3})=0$ which
represents a compact 3-dimensional object. In \fig{brill} the
2-dimensional intersection with the plane $\hat{k}_{3}=0$ is plotted
for several values of $c_{1}$ corresponding to different actions.\\ 
For our numerical studies, we discretized the Brillouin zone in finite  
intervals that could be made arbitrarily small. 

\begin{figure}
\begin{center}
\includegraphics[width=9cm]{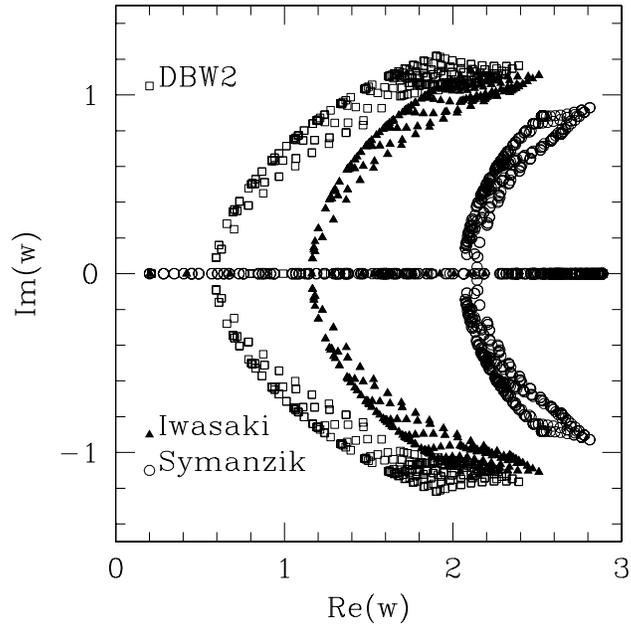}
\end{center}
\vspace{-1cm}
\caption{\footnotesize{\label{distr_poli}Distribution of the poles for different actions, scanning the Brillouin zone.}}
\end{figure}
In the region enclosed by the curve one has $\mathfrak{Im}(w)=0$, while outside $\mathfrak{Im}(w)\neq 0$.\\ 
The \fig{distr_poli} represents the distribution of the poles; by scanning the Brillouin zone and varying the intervals of the momenta,
one reaches the conclusion that
there exists a maximal value of $\mathfrak{Re}(w)$, which is related to $c_{1}$ by                                    
$$
w_{max}=2\textrm{arcsinh} \left(\frac{1}{2\sqrt{-2c_{1}}}\right)=
\left\{\begin{array}{ll}
 2.063 & \textrm{Symanzik, tree level}\\
 1.162 & \textrm{Iwasaki}\\
 0.588 & \textrm{DBW2}
\end{array}\right.
$$
such that for $w<w_{max}$
the imaginary part $\mathfrak{Im}(w)$ vanishes.\\
The value of $w_{max}$ yields an estimate of $\epsilon$ defined above through the simple relation\\
$$
\epsilon=e^{-w_{max}}=\left\{\begin{array}{ll}
0.127  & \textrm{Symanzik, tree level}\\
0.313  & \textrm{Iwasaki}\\
0.555  & \textrm{DBW2}
\end{array}\right.
$$
Evaluating observables from exponential decays of correlation
functions, one expects the presence of unphysical states unless $t\gg t_{min}$, where $t_{min}$ in lattice units is given by
\begin{equation}\label{tmin}
t_{min}=\frac{1}{w_{max}}=\left\{\begin{array}{ll}
 0.484 & \textrm{Symanzik, tree level}\\
 0.860 & \textrm{Iwasaki}\\
 1.702 & \textrm{DBW2}
\end{array}\right.
\end{equation}
One has to remember that our estimation of the unphysical poles has been performed in perturbation theory and hence has to be considered as an indicative evaluation.


%% file: results.tex
\section{Numerical results}\label{numerical_results}
The tables \ref{tab_iwasaki_results} and \ref{tab_dbw2_results} report the
data for the potential and the force evaluated at finite lattice spacing with
Iwasaki and DBW2 actions.
$\rI$ defines the tree-level improved force and is defined in \eq{e_rI_new}.
For the potential only the naive definition is reported.
\input tab_iwasaki.tex
\input tab_dbw2.tex

%% file: tab_iwasaki.tex
\begin{table}
\caption[The potential and the force in lattice units for the Iwasaki action.]{\footnotesize{The potential and the force in lattice units for the Iwasaki action.}\label{tab_iwasaki_results}}
\begin{center}
\begin{tabular}{c c c c c}
\hline
$\beta$   &  $r_{I}/a$ & $a^{2}F(r_{I})$ & $r/a$   &  $aV(r)$  \\
\hline
2.1551    &   1.4858 & 0.4059(21)  &  2        &  0.9519(20) \\  
          &   2.4626 & 0.2965(75)  &  3        &  1.2487(61)\\
          &   3.5663 & 0.2585(98)   &  4              &  1.515(12)\\
          &   4.6059 & 0.249(29)    &  5        &  1.783(12)\\
\hline
2.2879   & 1.4858 & 0.6055(29)   &  2   &     0.7730(12 \\
         & 2.4626 & 0.19796(76)  &  3   &     0.9707(33)\\
         & 3.5663 & 0.1689(70)    &  4  &     1.1418(41)\\
         & 4.6059 & 0.1622(64)   &  5   &     1.3047(95)\\
         & 5.6082 & 0.155(10)   &  6    &     1.460(16)\\
         & 6.6036 & 0.148(16)  &  7     &     1.608(33)\\
\hline
2.5208    & 1.4858 & 0.21301(25) &  2  &   0.60342(29) \\
          & 2.4626 & 0.11734(39) &  3  &   0.72099(46)\\
          & 3.5663 & 0.09011(70)  &  4  &   0.8107(14) \\
          & 4.6059 & 0.0803(10) &  5    &   0.8902(20)\\
          & 5.6082 & 0.07575(61)  &  6  &   0.9658(36)\\
          & 6.6036 & 0.0744(32)  &  7   &   1.0378(37)\\
          & 7.5977 & 0.0711(15)  &  8   &   1.1098(74) \\
\hline
2.7124    & 1.4858  &  0.17512(12)  &  2  &  0.52165(14) \\
          & 2.4626  & 0.08691(14)  &  3  &  0.60856(20) \\
          & 3.5663  & 0.06138(53) &  4  &  0.66994(56) \\
          & 4.6059   & 0.05194(81) &  5  &  0.7219(13) \\
          & 5.6082   & 0.04693(85)  &  6  &  0.7688(20) \\
          & 6.6036   &  0.0451(23) &  7  &  0.8139(11)  \\
          & 7.5977   & 0.0425(10)   &  8  &  0.8564(10)\\
          & 8.5915   & 0.0420(15) &  9  &  0.8984(23) \\
          & 9.5854   & 0.0414(19)  &  10 &  0.9398(25) \\
\hline
\end{tabular}
\end{center}
\end{table}


%% file: tab_dbw2.tex
\begin{table}
\caption[The potential and the force in lattice units for the DBW2 action.]{\footnotesize{The potential and the force in lattice units for the DBW2 action.}\label{tab_dbw2_results}}
\begin{center}
\begin{tabular}{c c c c c}
\hline
$\beta$   &  $r_{I}$ & $a^{2}F(r_{I})$ &  $r/a$   &  $aV(r)$  \\
\hline 
0.75696   &  1.9233  & 0.3885(13)  & 2 & 0.9004(17) \\ 
	  &  2.7793  &  0.2723(59) & 3 & 1.1726(33) \\ 
          &  3.7435  & 0.241(10)  & 4 & 1.410(11)  \\ 
          &  4.7364  & 0.2424(83)  & 5 & 1.658(20) \\ 
          &  5.7535  &  0.236(20)  & 6 & 1.894(35) \\ 
\hline
0.8243     & 1.9233 &  0.29697(61) & 2	&    0.7357(11) \\
           & 2.7793 & 0.1887(12)  & 3	&    0.9244(11)\\
	   & 3.7435 &  0.1613(12)  &  4	&    1.0857(17)  \\
           & 4.7364 &  0.1532(62)   &  5	&    1.2389(70)\\
	   &  5.7535 & 0.1476(91)  &  6	&    1.386(16)\\
	   & 6.7843 & 0.1494(61)  &  7	&    1.536(20)\\
\hline
0.9636    & 1.9233 & 0.20810(27)  &  2  &    0.55693(64)   \\     
          & 2.7793 &  0.1136(16) &  3  &    0.6703(20)  \\
          & 3.7435 & 0.08680(85) &  4	&    0.7569(20) \\
          & 4.7364 &  0.0783(13)  &  5	&    0.8340(29) \\
          & 5.7535 & 0.0745(37)  &  6	&    0.9076(57)\\
          & 6.7843 & 0.0700(57)  &  7	&    0.977(10)\\
          & 7.8120 & 0.0640(47)  &  8	&    1.038(10)\\
\hline
1.04      & 1.9233 & 0.18182(26) &  2  & 0.49710(24)  \\
          & 2.7793 & 0.09392(54)  &  3  & 0.59094(58) \\
          & 3.7435 & 0.06847(84)   &  4  & 0.65914(76) \\
          & 4.7364 & 0.0575(18)   &  5  & 0.7180(26)\\
          & 5.7535 & 0.0544(12)  &  6  & 0.7724(21) \\
          & 6.7843 & 0.0527(22)  &  7  & 0.8254(42) \\
          & 7.8120 & 0.0498(23)    &  8  & 0.8753(65) \\
          & 8.8296 & 0.0481(42)  &  9  & 0.9230(51)\\
          & 9.8369 & 0.0485(44)  &  10 & 0.9713(56) \\
          & 10.8365  & 0.0464(34) &  11 & 1.0175(95)  \\
	  & 11.8310  &  0.048(13) &  12 & 1.067(20)  \\
\hline
\end{tabular}
\end{center}
\end{table}

%% file: pap.bbl
\begin{thebibliography}{10}

\bibitem{Orginos:2001xa}
RBC, K. Orginos,
\newblock Nucl. Phys. Proc. Suppl. 106 (2002) 721, hep-lat/0110074,
\newblock 

\bibitem{Aoki:2002vt}
Y. Aoki et~al.,
\newblock (2002), hep-lat/0211023,
\newblock 

\bibitem{Jansen:2003jq}
K. Jansen and K.i. Nagai,
\newblock (2003), hep-lat/0305009,
\newblock 

\bibitem{Okamoto:1999hi}
CP-PACS, M. Okamoto et~al.,
\newblock Phys. Rev. D60 (1999) 094510, hep-lat/9905005,
\newblock 

\bibitem{pot:r0}
R. Sommer,
\newblock Nucl. Phys. B411 (1994) 839, hep-lat/9310022.

\bibitem{Iwasaki:1983ck}
Y. Iwasaki,
\newblock UTHEP-118.

\bibitem{deForcrand:1997bx}
QCD-TARO, P. de~Forcrand et~al.,
\newblock Nucl. Phys. Proc. Suppl. 53 (1997) 938, hep-lat/9608094,
\newblock 

\bibitem{Takaishi:1996xj}
T. Takaishi,
\newblock Phys. Rev. D54 (1996) 1050,
\newblock 

\bibitem{Weisz:1983zw}
P. Weisz,
\newblock Nucl. Phys. B212 (1983) 1,
\newblock 

\bibitem{Weisz:1984bn}
P. Weisz and R. Wohlert,
\newblock Nucl. Phys. B236 (1984) 397.

\bibitem{Hasenfratz:1994sp}
P. Hasenfratz and F. Niedermayer,
\newblock Nucl. Phys. B414 (1994) 785, hep-lat/9308004,
\newblock 

\bibitem{Niedermayer:2000yx}
F. Niedermayer, P. Rufenacht and U. Wenger,
\newblock Nucl. Phys. B597 (2001) 413, hep-lat/0007007,
\newblock 

\bibitem{Polyakov:1978vu}
A.M. Polyakov,
\newblock Phys. Lett. B72 (1978) 477,
\newblock 

\bibitem{Susskind:1979up}
L. Susskind,
\newblock Phys. Rev. D20 (1979) 2610,
\newblock 

\bibitem{Boyd:1996bx}
G. Boyd et~al.,
\newblock Nucl. Phys. B469 (1996) 419, hep-lat/9602007,
\newblock 

\bibitem{Beinlich:1997ia}
B. Beinlich, F. Karsch, E. Laermann and A. Peikert,
\newblock Eur. Phys. J. C6 (1999) 133, hep-lat/9707023,
\newblock 

\bibitem{deForcrand:1999bi}
QCD-TARO, P. de~Forcrand et~al.,
\newblock Nucl. Phys. B577 (2000) 263, hep-lat/9911033,
\newblock 

\bibitem{Bliss:1996wy}
D.W. Bliss, K. Hornbostel and G.P. Lepage,
\newblock (1996), hep-lat/9605041,
\newblock 

\bibitem{glueb:teper98}
M.J. Teper,
\newblock (1998), hep-th/9812187,
\newblock 

\bibitem{pot:r0_SU3}
ALPHA, M. Guagnelli, R. Sommer and H. Wittig,
\newblock Nucl. Phys. B535 (1998) 389, hep-lat/9806005,
\newblock 

\bibitem{Necco:2001xg}
S. Necco and R. Sommer,
\newblock Nucl. Phys. B622 (2002) 328, hep-lat/0108008,
\newblock 

\bibitem{smear:ape}
APE, M. Albanese et~al.,
\newblock Phys. Lett. 192B (1987) 163.

\bibitem{PPR}
G. Parisi, R. Petronzio and F. Rapuano,
\newblock Phys. Lett. 128B (1983) 418.

\bibitem{HOR1}
M. Creutz,
\newblock Phys. Rev. D36 (1987) 515.

\bibitem{HOR2}
F.R. Brown and T.J. Woch,
\newblock Phys. Rev. Lett. 58 (1987) 2394.

\bibitem{pot:r0_SCRI}
R.G. Edwards, U.M. Heller and T.R. Klassen,
\newblock (1997), hep-lat/9711003.

\bibitem{Luscher:1984is}
M. Luscher and P. Weisz,
\newblock Nucl. Phys. B240 (1984) 349,
\newblock 

\bibitem{varia:michael}
N.A. Campbell, A. Huntley and C. Michael,
\newblock Nucl. Phys. B306 (1988) 51.

\bibitem{phaseshifts:LW}
M. {L\"uscher} and U. Wolff,
\newblock Nucl. Phys. B339 (1990) 222.

\bibitem{Necco:2001gh}
S. Necco and R. Sommer,
\newblock Phys. Lett. B523 (2001) 135, hep-ph/0109093,
\newblock 

\bibitem{billoire}
B. Berg and A. Billoire,
\newblock Nucl. Phys. B221 (1983) 109,
\newblock 

\bibitem{wenger}
U. Wenger,
\newblock (2000), Ph.D.Thesis, unpub.,
\newblock 

\bibitem{Ishikawa:1983xg}
K. Ishikawa, G. Schierholz and M. Teper,
\newblock Z. Phys. C19 (1983) 327,
\newblock 

\bibitem{Vaccarino:1999ku}
A. Vaccarino and D. Weingarten,
\newblock Phys. Rev. D60 (1999) 114501, hep-lat/9910007,
\newblock 

\bibitem{Bali:1993fb}
UKQCD, G.S. Bali et~al.,
\newblock Phys. Lett. B309 (1993) 378, hep-lat/9304012,
\newblock 

\bibitem{Wittig:1999kb}
H. Wittig,
\newblock (1999), hep-ph/9911400,
\newblock 

\bibitem{Morningstar:1999rf}
C.J. Morningstar and M.J. Peardon,
\newblock Phys. Rev. D60 (1999) 034509, hep-lat/9901004,
\newblock 

\bibitem{Luscher:2001up}
M. Luscher and P. Weisz,
\newblock JHEP 09 (2001) 010, hep-lat/0108014,
\newblock 

\bibitem{Meyer:2002cd}
H.B. Meyer,
\newblock JHEP 01 (2003) 048, hep-lat/0209145,
\newblock 

\bibitem{Blum:1995xb}
T. Blum et~al.,
\newblock Nucl. Phys. B442 (1995) 301, hep-lat/9412038,
\newblock 

\bibitem{Aoki:2001xq}
JLQCD, S. Aoki et~al.,
\newblock Nucl. Phys. Proc. Suppl. 106 (2002) 263, hep-lat/0110088,
\newblock 

\end{thebibliography}
